\documentclass[lettersize,journal]{IEEEtran}
\IEEEoverridecommandlockouts

\usepackage{cite}
\usepackage{amsmath,amssymb,amsfonts}
\usepackage[protrusion=true,expansion=true]{microtype}
\usepackage{algorithmic}
\usepackage{algorithm}
\usepackage{array}
\usepackage[caption=false,font=normalsize,labelfont=sf,textfont=sf]{subfig}
\usepackage{textcomp}
\usepackage{url}
\usepackage{graphicx}
\usepackage{xcolor}
\usepackage{booktabs}
\usepackage{multirow}
\usepackage[hidelinks]{hyperref}
\usepackage{tikz}
\usetikzlibrary{shapes.geometric, arrows.meta, calc, positioning, decorations.pathmorphing, decorations.pathreplacing, backgrounds, fit, patterns, shadows}
\usepackage{pgfplots}
\pgfplotsset{compat=1.18}
\usepgfplotslibrary{groupplots}
\usepackage{balance}

\definecolor{c32}{HTML}{1F77B4}    
\definecolor{cf32}{HTML}{D62728}   
\definecolor{cckks}{HTML}{2CA02C}  
\definecolor{cgridln}{HTML}{DCDCDC}
\definecolor{cdeg2}{HTML}{E07B39}  
\definecolor{cdeg1}{HTML}{2CA02C}  

\pgfplotsset{
  faultbase/.style={
    width=0.48\textwidth,
    height=4.3cm,
    xmode=log,
    log basis x=10,
    xlabel={Bit Error Rate (BER)},
    xmin=5e-8, xmax=2e-3,
    grid=none,
    ymajorgrids=true,
    y grid style={line width=0.3pt, draw=cgridln},
    minor tick num=0,
    tick align=outside,
    tickpos=left,
    axis line style={draw=black!55, line width=0.5pt},
    every axis title/.append style={font=\small\bfseries, yshift=1pt},
    tick label style={font=\scriptsize},
    label style={font=\small},
    legend cell align=left,
    legend style={
      font=\scriptsize,
      draw=black!25,
      fill=white,
      fill opacity=1,
      text opacity=1,
      rounded corners=1.5pt,
      inner sep=2.5pt,
      row sep=0.5pt,
    },
  },
  int32line/.style={c32, very thick, mark=square*, mark size=2pt,
      mark options={fill=c32, draw=white, line width=0.4pt}},
  float32line/.style={cf32, very thick, densely dashed, mark=triangle*, mark size=2.6pt,
      mark options={solid, fill=cf32, draw=white, line width=0.4pt}},
  ckksline/.style={cckks, very thick, dashdotted, mark=diamond*, mark size=2.8pt,
      mark options={solid, fill=cckks, draw=white, line width=0.4pt}},
  randline/.style={black!55, thin, densely dotted, forget plot},
  band32/.style={c32, fill opacity=0.12, draw=none, forget plot},
  ebcfg/.style={error bars/.cd, y dir=both, y explicit,
      error bar style={line width=0.6pt, draw=c32!85!black},
      error mark=-, error mark options={c32!85!black, line width=0.6pt, mark size=2.4pt}},
}

\def\BibTeX{{\rm B\kern-.05em{\sc i\kern-.025em b}\kern-.08em
    T\kern-.1667em\lower.7ex\hbox{E}\kern-.125emX}}
\usepackage{fancyhdr}

\fancypagestyle{firstpage}
{
    \fancyhead[L]{\footnotesize Personal use of this material is permitted.  Permission must be obtained for all other uses, in any current or future media, including reprinting/republishing this material for advertising or promotional purposes, creating new collective works, for resale or redistribution to servers or lists, or reuse of any copyrighted component of this work in other works.This manuscript has been submitted to an IEEE Transactions journal and is currently under review.}
    \fancyhead[R]{}
}
\begin{document}

\title{PRISM: \textbf{\underline{S}}ensitivity-Aware \textbf{\underline{P}}olyno\textbf{\underline{M}}ial P\textbf{\underline{R}}uning for Eff\textbf{\underline{I}}cient Neural Network Encryption}

\author{Sahaj~Majavdia and Mahdi~Taheri%
\thanks{Sahaj Majavdia is with Brandenburgische Technische Universität Cottbus-Senftenberg, Germany. Mahdi Taheri is with Humboldt University of Berlin, Germany and also with Tallinn University of Technology, Tallinn, Estonia (e-mail: sahaj.majavdia@b-tu.de; mahdi.taheri@taltech.ee).}}


\maketitle
\thispagestyle{firstpage}

\begin{abstract}
Structured pruning is essential for making neural network inference
feasible under homomorphic encryption (HE), yet its impact on model
reliability has remained unexplored. This paper presents a systematic
reliability characterization of pruned CKKS-encrypted neural networks
and introduces Polynomial-Sensitivity-Aware Pruning (PSAP), a
structured pruning method that is inherently reliability-aware. PSAP
scores filters jointly by weight magnitude, polynomial activation
sensitivity, and rotation cost, which concentrates pruning in
fault-tolerant regions. Across two architectures, two datasets, two
numerical representations, and five bit-error rates (40 full-model and
108 per-layer experiments), PSAP-pruned models limit catastrophic
($>$10\,pp accuracy drop) layers to at most two versus 5--14 for
magnitude-pruned baselines, reducing worst-case vulnerability by up to 29$\times$ (1.4$\times$--29$\times$ across configurations) under int32
bit-flip injection. Direct CKKS encrypted fault injection indicates a
safe operating boundary near BER~$\leq 10^{-5}$, supporting int32
injection as a conservative reliability proxy. The fault-critical
structural layers account for only $1.1\%$ of parameters, enabling
selective hardening at minimal overhead. These reliability gains are
obtained alongside competitive efficiency: PSAP reduces
Halevi--Shoup rotations by up to 45.2\% on ResNet-32, and an
adaptive mixed-degree allocation scheme lowers multiplicative depth
from 66 to 56 levels, enabling leveled inference without bootstrapping.
\end{abstract}


\section{Introduction}
\label{sec:introduction}

\IEEEPARstart{H}{omomorphic} encryption (HE), particularly the
Cheon--Kim--Kim--Song (CKKS) scheme~\cite{cheon2017ckks}, enables
neural network inference directly on encrypted data, making it an
attractive solution for privacy-sensitive applications such as
healthcare diagnostics, financial fraud detection, and biometric
authentication. Since plaintext inputs are never exposed during
computation, inference can be offloaded to untrusted cloud servers
while preserving data confidentiality. However, this privacy comes
at a significant computational cost. Each ciphertext consists of
polynomial coefficients represented under a large coefficient
modulus $Q$ (1260--1760~bits in typical configurations), and every
homomorphic operation requires computationally expensive
Halevi--Shoup rotations~\cite{halevi2014algorithms}. Consequently,
structured pruning has become an important optimization technique
for reducing the computational overhead of HE inference.

Several HE-aware pruning methods have been proposed to reduce this
computational overhead. Hunter~\cite{cai2022hunter} identifies HE-friendly structures for pruning, SpENCNN~\cite{ran2023spencnn}
jointly optimizes single instruction multiple data (SIMD) encoding and sub-block sparsity, MOSAIC~\cite{cai2024mosaic} adopts a
prune-and-assemble strategy, and PrivCirNet~\cite{xu2024privcirnet}
employs block circulant transformations. These methods focus
primarily on improving computational efficiency by reducing rotation counts and optimizing ciphertext packing. However, the pruning process is guided by conventional plaintext criteria, such as weight magnitude or structural alignment, without considering the sensitivity of the polynomial activation functions used in HE inference.

Reliability under hardware faults is a critical concern for
safety-critical applications deployed on cloud and edge platforms.
Transient faults, such as cosmic-ray-induced single-event
upsets~\cite{mukherjee2005soft,baumann2005radiation}, can corrupt
bits stored in dynamic random access memory (DRAM), caches, or
accelerator registers. The impact of such faults on plaintext neural networks has been extensively investigated~\cite{reagen2018ares,li2017understanding, mahmoud2021optimizing,chen2019binfi}.

Compared to conventional plaintext inference, homomorphic encryption (HE) inference places substantially greater demands on the underlying computing system. The large ciphertext representation significantly increases the memory footprint, while homomorphic operations incur considerably longer execution times. Together, these characteristics increase the exposure of encrypted inference to transient hardware faults by enlarging both the amount of memory susceptible to bit corruptions and the duration over which such faults may occur. Consequently, random memory bit-flips are more likely to affect HE workloads than conventional neural network inference. Moreover, faults introduced into ciphertext coefficients propagate through subsequent homomorphic operations, potentially corrupting the entire encrypted computation and resulting in silent data corruption without any indication to the user. Recent work has begun to investigate the reliability of fully homomorphic encryption (FHE) at the cryptographic-operation level~\cite{rajagede2025fhereliability,mu2026fhesdc}, showing that a single bit-flip in a ciphertext polynomial can invalidate an entire homomorphic computation; these efforts are discussed in detail in Section~\ref{sec:related}.

However, these studies focus on the reliability of cryptographic operations and hardware implementations without considering the effect of model optimization on the reliability of encrypted inference. Structured pruning and polynomial approximation directly modify the computational graph, the distribution of model parameters, and the sequence of homomorphic operations, all of which can influence fault propagation and model resilience. Despite their importance for efficient HE inference, the reliability implications of these optimization techniques remain largely unexplored. To address this gap, this paper presents Polynomial-Sensitivity-Aware Pruning (PSAP), a reliability-aware structured pruning framework for CKKS-encrypted neural networks. Unlike existing HE-aware pruning approaches that primarily optimize computational efficiency, PSAP jointly considers filter importance, polynomial activation sensitivity, and homomorphic rotation cost to guide pruning toward more fault-tolerant network structures. Furthermore, an adaptive mixed-degree polynomial allocation strategy reduces the multiplicative depth required for encrypted inference, while a comprehensive fault injection framework evaluates the reliability impact of each optimization decision under transient hardware faults. The main contributions of this paper are summarized as follows: 
\begin{itemize} 
\item A systematic reliability characterization framework for pruned CKKS-encrypted neural networks under transient bit-flip faults, enabling the analysis of fault propagation across different network architectures, datasets, numerical representations, and fault conditions.

\item A reliability-aware structured pruning framework, Polynomial-Sensitivity-Aware Pruning (PSAP), that jointly considers weight importance, polynomial activation sensitivity, and homomorphic rotation cost to optimize both fault tolerance and computational efficiency.

\item A comprehensive fault analysis methodology for encrypted neural networks, combining plaintext and direct CKKS fault injection to investigate fault propagation, identify fault-critical components, and evaluate the reliability of encrypted inference.

\item An adaptive mixed-degree polynomial allocation strategy for CKKS-encrypted neural networks that jointly optimizes multiplicative depth, computational efficiency, and inference feasibility without requiring bootstrapping.
\end{itemize}

\section{Related Work}
\label{sec:related}

\subsection*{HE-Aware Neural Network Optimization}

CryptoNets~\cite{gilad2016cryptonets} first demonstrated neural inference on
encrypted data. GAZELLE~\cite{juvekar2018gazelle} introduced a hybrid
HE/garbled-circuit approach, and CryptoNAS~\cite{ghodsi2020cryptonas} further optimized the network architecture to minimize the ReLU budget for such hybrid protocols. Lee et al.~\cite{lee2022heresnet} scaled CKKS to
ResNet architectures. HyPHEN~\cite{kim2024hyphen} achieves 1.4\,s on GPU via
rotation-free aggregation. These methods focus on enabling or accelerating
HE inference but do not address the pruning criterion itself.

Hunter~\cite{cai2022hunter} identifies HE-friendly structures to prune
operations aligned with ciphertext packing, achieving 49\% permutation
reduction on ResNet-32. MOSAIC~\cite{cai2024mosaic} extends this with a
prune-and-assemble strategy, reducing computational cost by up to 91\% on
ResNet-50. SpENCNN~\cite{ran2023spencnn} co-designs SIMD encoding with
sub-block weight pruning, reporting 1.87$\times$ speedup on ResNet-20.
PrivCirNet~\cite{xu2024privcirnet} transforms weights into block circulant
matrices, achieving 2--7$\times$ latency reduction (NeurIPS~2024).
HE-PEx~\cite{aharoni2023hepex} uses tile-tensor permutations, achieving
10--35\% latency reduction, while MOFHEI~\cite{mofhei2024} systematically optimizes the model architecture for efficient HE execution. All these methods optimize sparsity patterns or
weight structure for ciphertext layout but select pruning targets using
standard plaintext heuristics (magnitude, group lasso, or structural
alignment); none accounts for the sensitivity profile of the polynomial
activations that replace ReLU in HE inference, nor has any prior
HE-aware pruning method evaluated the fault tolerance of the pruned
models.

AutoFHE~\cite{ao2024autofhe} automates degree selection but does not prune
filters. Standard magnitude ($\ell_1$-norm~\cite{li2017pruning}) and
reconstruction-error~\cite{he2017channel} pruning reduce FLOPs but ignore
rotation count, the dominant HE cost, and polynomial sensitivity. A unified
criterion that jointly accounts for rotation cost, polynomial sensitivity,
and weight magnitude remains an open problem.

\subsection*{Fault Tolerance in Encrypted Computation}

The interaction between hardware faults and homomorphic encryption has been studied across distinct layers of the system stack. Earlier efforts in FHE reliability focused on cryptographic noise management: for instance, Chillotti et al.~\cite{chillotti2020tfhe} utilize TFHE gate bootstrapping to refresh ciphertexts, though this addresses internal cryptographic noise accumulation rather than physical hardware faults, and is limited to exact Boolean logic rather than approximate arithmetic schemes like CKKS.

At the hardware and cryptographic layers, physical memory faults have also been characterized. Rajagede and Solihin~\cite{rajagede2025fhereliability} analyze the propagation of memory faults through core FHE primitive operations (NTT and RNS decomposition), demonstrating that a single bit-flip in a ciphertext polynomial can invalidate the entire homomorphic evaluation. In a parallel study, Mu et al.~\cite{mu2026fhesdc} provide the first comprehensive characterization of Silent Data Corruptions (SDCs) in CKKS-encrypted operations, demonstrating that single-bit flips in ciphertexts lead to a $22\%$ SDC rate. They show that slot-level error magnitudes scale monotonically with the ciphertext modulus $Q$ (ranging up to $10^{125}$), multiplication and rescaling operations damp error growth, and conventional hardware/software mitigations such as Dual Modular Redundancy (DMR) and checksum-based Algorithm-Based Fault Tolerance (ABFT) introduce significant latency overheads ($1.15\times\text{--}2.0\times$). From a security perspective, Mankali et al.~\cite{mankali2025glitchfhe} (GlitchFHE) demonstrate adversarial fault injection attacks on physical FHE accelerators.

These lines of work occupy different regions of the design space than the present study, representing a difference in scope rather than performance. Traditional DNN reliability studies~\cite{reagen2018ares,li2017understanding,chen2019binfi,mahmoud2021optimizing} characterize fault propagation and selective protection (such as selective hardening or TMR) in unencrypted models but do not reach the encrypted execution domain. FHE reliability studies~\cite{rajagede2025fhereliability,mankali2025glitchfhe,mu2026fhesdc} operate purely at the low-level cryptographic operator or hardware accelerator level, without an algorithmic, model-level, or layer-level view of the neural network. No prior work investigates how model optimization decisions (such as structured pruning or mixed-degree activation allocation) reshape the layer-wise fault tolerance of CKKS-encrypted networks. Whereas these operator-level studies characterize how individual cryptographic primitives fail, the evaluation in this paper targets the orthogonal dimension left open by both, providing the first model-level characterization of layer-wise vulnerability and establishing empirical safe operating boundaries for encrypted neural network inference.

\begin{figure*}[!t]
\centering
\scalebox{0.72}{%
\begin{tikzpicture}[
  font=\sffamily,
  >=Latex,
  phase/.style={rectangle, draw=black!60, fill=black!4, rounded corners=2pt,
    minimum width=2.2cm, minimum height=0.9cm, align=center,
    line width=0.6pt, inner sep=2pt, drop shadow={opacity=0.05, shadow xshift=1pt, shadow yshift=-1pt}},
  state/.style={font=\scriptsize\sffamily\itshape, text=black!80, align=center},
  pnote/.style={font=\scriptsize\sffamily, text=black!60, align=center},
  arr/.style={->, line width=0.8pt, draw=black!60},
  darr/.style={->, line width=0.8pt, draw=black!40, densely dashed},
  glabel/.style={font=\scriptsize\sffamily\bfseries, text=black!50}
]
\node[state, align=right] (in) {\textbf{Plain CNN}\\ReLU + BN};
\node[phase, right=1.0cm of in] (p1) {{\small\bfseries Phase 1}\\[-1pt]{\scriptsize Poly Conversion}};
\node[pnote, below=2pt of p1] (n1) {ReLU to Polynomial\\[1pt]BN Folding};
\node[phase, right=1.8cm of p1] (p2) {{\small\bfseries Phase 2}\\[-1pt]{\scriptsize PSAP Pruning}};
\node[pnote, below=2pt of p2] (n2) {Sensitivity-Aware\\[1pt]Rotation-Aware\\[1pt]Reliability-Aware};
\node[phase, right=1.8cm of p2] (p3) {{\small\bfseries Phase 3}\\[-1pt]{\scriptsize Mixed-Degree}};
\node[pnote, below=2pt of p3] (n3) {Deg-2/Deg-1 Assign.\\[1pt]Depth Reduction};
\node[phase, right=1.8cm of p3] (p4) {{\small\bfseries Phase 4}\\[-1pt]{\scriptsize QAT + Export}};
\node[pnote, below=2pt of p4] (n4) {Quant.-Aware Training\\[1pt]HE Model Export};
\node[state, right=1.0cm of p4, align=left] (out) {\textbf{HE Ready}\\CNN};

\node[rectangle, draw=red!60, fill=red!4, rounded corners=2pt,
      minimum width=2.4cm, minimum height=0.9cm, align=center,
      line width=0.6pt, right=1.0cm of out,
      drop shadow={opacity=0.05, shadow xshift=1pt, shadow yshift=-1pt}] (rel)
      {{\small\bfseries Reliability}\\[-1pt]{\scriptsize Evaluation}};
\node[pnote, below=2pt of rel] (nrel) {Bit-Flip Injection\\[1pt]Per-Layer Analysis};
\draw[arr] (in) -- (p1);
\draw[arr] (p1) -- node[above, state] {Poly.\\CNN} (p2);
\draw[arr] (p2) -- node[above, state] {Sparse\\CNN} (p3);
\draw[arr] (p3) -- node[above, state] {Depth-Opt.\\CNN} (p4);
\draw[arr] (p4) -- (out);
\draw[arr] (out) -- (rel);
\coordinate (mid_pipe) at ($(p2)!0.5!(p3)$);
\node[rectangle, draw=black!60, fill=blue!4, rounded corners=3pt,
      minimum width=6.5cm, minimum height=1.3cm, align=center, line width=0.6pt,
      below=2.0cm of mid_pipe, drop shadow={opacity=0.05, shadow xshift=1pt, shadow yshift=-1pt}] (server)
      {{\small\bfseries Untrusted Server (OpenFHE CKKS)}\\[2pt]
       {\scriptsize Encrypted Inference: Conv to Poly Act to Rot \& Rescale}};

\node[rectangle, draw=black!60, fill=black!4, rounded corners=2pt,
      minimum width=2cm, minimum height=0.9cm, align=center, line width=0.6pt,
      left=1.5cm of server] (client_in) {{\small\bfseries Client}\\[-1pt]{\scriptsize Encrypts Image}};
\node[rectangle, draw=black!60, fill=black!4, rounded corners=2pt,
      minimum width=2cm, minimum height=0.9cm, align=center, line width=0.6pt,
      right=1.5cm of server] (client_out) {{\small\bfseries Client}\\[-1pt]{\scriptsize Decrypts Pred.}};
\draw[arr] (client_in) -- node[above, font=\scriptsize\sffamily] {$\mathsf{Enc}(x)$}
                         node[below, font=\scriptsize\sffamily, text=black!60] (evalkeys) {+ Eval Keys} (server);
\draw[arr] (server) -- node[above, font=\scriptsize\sffamily] {$\mathsf{Enc}(y)$} (client_out);

\draw[darr, rounded corners=3pt] (out.south) -- ++(0, -1.2) -| node[pos=0.25, below, pnote] {Deploy Model} (server.north);

\node[anchor=west, xshift=0.55cm] at (client_out.east) {\includegraphics[width=2.6cm]{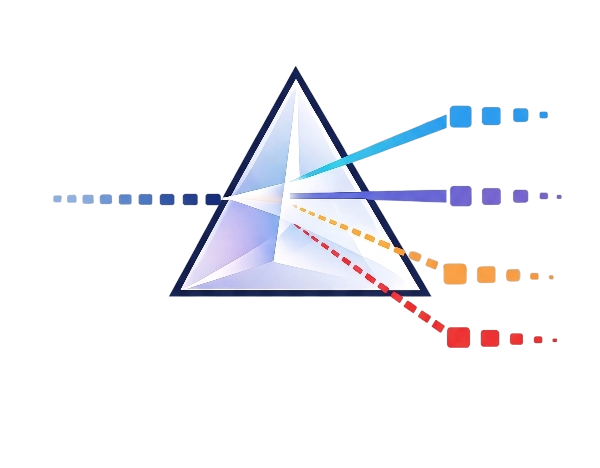}};
\end{tikzpicture}%
}
\caption{End-to-end PSAP pipeline: optimization phases (top), reliability evaluation (top-right), and deployed encrypted inference (bottom).}
\label{fig:psap_pipeline}
\end{figure*}
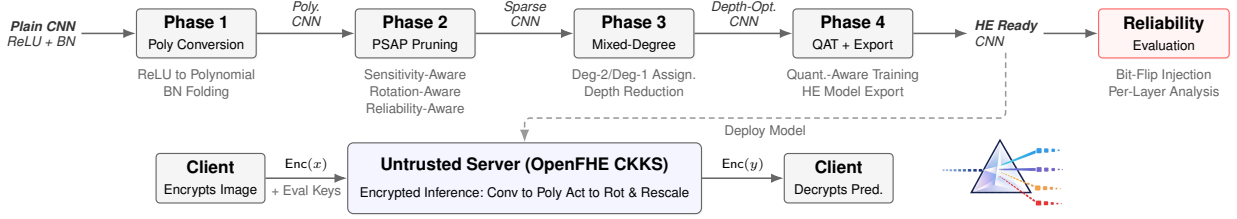

\section{Methodology}
\label{sec:method}

This section presents the proposed reliability-aware optimization
framework for homomorphic encryption (HE)-based neural network
inference. The framework jointly addresses the computational
constraints imposed by the Cheon--Kim--Kim--Song (CKKS) scheme and
the reliability challenges associated with transient hardware
faults. Rather than treating efficiency optimization and
reliability evaluation as independent processes, the proposed
methodology integrates both objectives into a unified optimization
flow that transforms a pretrained convolutional neural network into
an HE-compatible model suitable for efficient and reliable
encrypted inference.

As illustrated in Fig.~\ref{fig:psap_pipeline}, the proposed
framework consists of four sequential optimization stages. The
network is first transformed into an HE-compatible representation
through polynomial activation modeling. The transformed model is
then optimized using the proposed
Polynomial-Sensitivity-Aware Pruning (PSAP), which exploits
activation sensitivity, filter importance, and homomorphic
computation cost to guide structured pruning. Next, an adaptive
mixed-degree polynomial allocation strategy reduces the
multiplicative depth required for encrypted inference while
preserving the nonlinear behavior of the most influential
activations. Finally, quantization-aware fine tuning generates the
model deployed within the OpenFHE CKKS inference pipeline. The
resulting network is subsequently characterized under transient
memory faults using the proposed reliability modeling framework.

Unlike conventional HE optimization methods that focus solely on
reducing computational complexity, the proposed methodology jointly
optimizes computational efficiency, multiplicative depth, and
intrinsic fault tolerance. The optimization therefore aims not only
to reduce the execution cost of encrypted inference, but also to
preserve network structures that contribute to reliable operation
under transient hardware faults. This flow is complemented by a
reliability modeling framework that relates the pruning and
degree-allocation decisions to their impact on fault propagation and
layer vulnerability, so that efficiency and fault tolerance are
analyzed within the same framework.

\subsection{Overall Framework}

Let

\begin{equation}
\mathcal{M}_0=
\left\{
W_1,\ldots,W_M,
\phi_1,\ldots,\phi_M
\right\}
\label{eq:model_def}
\end{equation}

denote a pretrained convolutional neural network consisting of
weight tensors $W_l$ and activation functions $\phi_l$.

The proposed methodology transforms the initial network into an
optimized model through a sequence of optimization operators,

\begin{equation}
\mathcal{M}^{*}
=
\mathcal{T}_{\mathrm{QAT}}
\circ
\mathcal{T}_{\mathrm{Deg}}
\circ
\mathcal{T}_{\mathrm{PSAP}}
\circ
\mathcal{T}_{\mathrm{Poly}}
\left(
\mathcal{M}_0
\right),
\label{eq:framework}
\end{equation}

where
$\mathcal{T}_{\mathrm{Poly}}$
converts the network into an HE-compatible architecture through
polynomial activation replacement,
$\mathcal{T}_{\mathrm{PSAP}}$
performs reliability-aware structured pruning,
$\mathcal{T}_{\mathrm{Deg}}$
optimizes the polynomial degrees to satisfy the available
multiplicative depth budget, and
$\mathcal{T}_{\mathrm{QAT}}$
denotes the quantization-aware fine-tuning stage that restores the
prediction accuracy of the optimized model before deployment.

Each transformation addresses a different optimization objective
while preserving the modifications introduced by the previous
stages. Consequently, the optimized model
$\mathcal{M}^{*}$
simultaneously satisfies the computational constraints of CKKS
inference and the reliability requirements considered throughout
this work. The optimized network is finally deployed within the
OpenFHE CKKS runtime and analyzed using the reliability modeling
framework presented later in this section.

\subsection{Design Objectives}

The proposed methodology simultaneously satisfies four complementary
objectives. First, the optimized model must remain compatible with
leveled CKKS inference, replacing all unsupported nonlinear operations
with polynomial approximations while keeping the total multiplicative
depth within the available level budget. Second, the encrypted-inference
overhead should be minimized by reducing homomorphic rotations,
ciphertext operations, and memory through HE-aware structured pruning.
Third, the optimization should preserve the intrinsic fault tolerance of
the network by avoiding pruning decisions that increase the vulnerability
of fault-critical layers or amplify fault propagation. Finally, the model
should maintain prediction accuracy while remaining deployable without
bootstrapping under the selected CKKS security parameters. The following
subsections describe how each stage contributes to these objectives.

\subsection{Polynomial Activation Modeling}
\label{sec:poly_model}

The first optimization stage transforms the pretrained neural
network into an architecture compatible with leveled homomorphic
encryption (HE) inference. Since the Cheon--Kim--Kim--Song (CKKS)
scheme supports only additions and multiplications, conventional
nonlinear activation functions containing comparison operations
cannot be directly evaluated on encrypted data. Consequently, all
unsupported nonlinear operators are replaced with trainable
polynomial approximations while preserving the predictive capability
of the original network. This transformation defines the operator
$\mathcal{T}_{\mathrm{Poly}}$ introduced in
Eq.~(\ref{eq:framework}) and provides the activation sensitivity
information required by the subsequent reliability-aware pruning
stage.

\subsubsection{Polynomial Activation Replacement}
\label{sec:poly_replace}

Recall the pretrained network
$\mathcal{M}_0$ from Eq.~(\ref{eq:model_def}), consisting of
weight tensors $W_l$ and activation functions $\phi_l$.

Each rectified linear unit (ReLU) activation is replaced by a
trainable polynomial using the Self-Learning Activation Function
(SLAF) formulation~\cite{pulido2024slaf}, extending earlier
polynomial-based HE activation
approaches~\cite{ishiyama2020poly},

\begin{equation}
p(x)
=
ax^2+bx+c,
\label{eq:poly_activation}
\end{equation}

where the coefficients $a$, $b$, and $c$ are optimized jointly with
the network parameters during retraining.

Degree-two polynomials are adopted because they provide the highest
nonlinearity that remains compatible with leveled CKKS inference
under the selected security parameters. Evaluating a polynomial of degree $d$
requires $d$ multiplicative levels. Increasing the polynomial degree
generally improves the approximation accuracy of ReLU, but also
increases the multiplicative depth of the encrypted computation,
eventually requiring bootstrapping, which remains a costly
operation even with state-of-the-art
algorithms~\cite{bossuat2021bootstrapping}. Degree-two activations
therefore provide an effective compromise between approximation
accuracy and computational feasibility.

To further simplify the encrypted computation graph, every batch
normalization layer is folded into its preceding convolution by
absorbing the normalization parameters into the convolution weights
and biases. The transformed network therefore consists only of
operations directly supported by the CKKS evaluation pipeline,
including convolutions, polynomial activations, residual additions,
average pooling, and fully connected layers.

Following the architectural transformation, the network is retrained
to recover the prediction accuracy lost during the replacement of
ReLU activations, producing the transformed model

\begin{equation}
\mathcal{M}_{\mathrm{Poly}}
=
\mathcal{T}_{\mathrm{Poly}}
(\mathcal{M}_0),
\end{equation}

which serves as the input to the subsequent optimization stages.

\subsubsection{Activation Sensitivity Characterization}
\label{sec:poly_sensitivity}

Replacing ReLU with trainable polynomial activations introduces
non-uniform activation behavior across different operating regions.
Unlike ReLU, whose derivative is piecewise constant, polynomial
activations exhibit continuously varying gradients. Consequently,
the contribution of each filter depends not only on its weights but
also on the local behavior of its activation function.

To quantify this behavior, activation statistics are collected for
every output channel using a representative calibration subset of
the training data. Let $\mu_i$ and $\sigma_i$ denote the mean and
standard deviation of the activation distribution corresponding to
filter $i$. The effective operating interval is approximated as

\begin{equation}
[\mu_i-3\sigma_i,\,
 \mu_i+3\sigma_i],
\end{equation}

which contains the majority of the observed activation values.

The activation sensitivity is defined as the expected magnitude of
the polynomial gradient over the corresponding operating interval, assuming a uniform input distribution,

\begin{equation}
S_i
=
\mathbb{E}_{x\in[\mu_i\pm3\sigma_i]}
\left[
\left|
\frac{\partial p(x)}
{\partial x}
\right|
\right],
\label{eq:poly_sensitivity}
\end{equation}

where, for the adopted quadratic polynomial,

\begin{equation}
\frac{\partial p(x)}
{\partial x}
=
2ax+b.
\label{eq:poly_derivative}
\end{equation}

The activation sensitivity $S_i$ provides a quantitative estimate
of the information preserved by each feature map after polynomial
transformation. Filters with large $S_i$ operate in steep polynomial
regions and therefore contribute more strongly to feature propagation,
whereas filters with small $S_i$ reside in flat regions where the
activation response varies little. Rather than directly determining
pruning decisions, this quantity serves as one of the optimization
variables employed by the proposed reliability-aware structured
pruning framework.

\subsection{Reliability-Aware Structured Pruning}
\label{sec:psap}

The transformed model
$\mathcal{M}_{\mathrm{Poly}}$
still contains substantial structural redundancy and is subsequently
optimized through the proposed Polynomial-Sensitivity-Aware Pruning
(PSAP), corresponding to the operator
$\mathcal{T}_{\mathrm{PSAP}}$
in Eq.~(\ref{eq:framework}).
Conventional structured pruning ranks filters by criteria such as
the $\ell_1$/$\ell_2$-norm or reconstruction error; although
effective for floating-point inference, these metrics do not reflect
the characteristics of encrypted execution. Instead, PSAP jointly
considers three complementary objectives during filter selection:
predictive importance, activation sensitivity (filters in highly
sensitive polynomial regions propagate more information), and
homomorphic evaluation cost (the Halevi--Shoup rotation count varies
with each layer's dimensions). The resulting optimization reduces
the computational overhead of encrypted inference while preserving
network structures that are expected to contribute most strongly to
reliable operation under transient hardware faults, avoiding
excessive pruning of structurally critical layers.

\subsubsection{PSAP Importance Metric}
\label{sec:psap_metric}

Consider convolution layer $\ell$ containing
$N_{\ell}$ output filters,

\begin{equation}
\mathcal{F}_{\ell}
=
\{
f_1,f_2,\ldots,f_{N_{\ell}}
\}.
\end{equation}

Each filter is characterized by three complementary optimization
variables.

The predictive importance is represented by the weight magnitude
$m_i = \|W_i\|_2$, where $W_i$ denotes the convolution kernel
associated with filter $i$.

The second variable is the polynomial activation sensitivity
$S_i$ obtained from
Eq.~(\ref{eq:poly_sensitivity}), which estimates the
contribution of the corresponding feature map after polynomial
transformation.

The third variable quantifies the computational benefit associated
with pruning layer $\ell$ through the normalized Halevi--Shoup
rotation cost $R_{\ell} = r_{\ell} / \max_j r_j$, where
$r_{\ell}$ is the number of homomorphic rotations required to
evaluate layer $\ell$.

Since the early feature extraction layers generally exhibit higher
information density, an additional depth-dependent protection factor
is introduced,

\begin{equation}
\Pi_{\ell}
=
1
+
2
(1-\delta_{\ell}),
\label{eq:depth_factor}
\end{equation}

where $\delta_{\ell}\in[0,1]$ denotes the normalized network depth. The
linear form is deliberately simple, granting the earliest layers
($\delta_{\ell}{=}0$) a threefold importance boost that decays to unity at
the output ($\delta_{\ell}{=}1$), reflecting the higher information density
and broader fault influence of early feature extraction stages; only
the relative ordering it induces affects pruning, so the specific
slope is not critical.

Combining these quantities yields the proposed PSAP importance
metric,

\begin{equation}
I_i
=
\frac
{
m_i
\,
S_i^{\alpha}
}
{
1+\lambda R_{\ell}
}
\Pi_{\ell},
\label{eq:psap}
\end{equation}

where $\alpha$ controls the contribution of activation sensitivity
and $\lambda$ determines the relative importance of homomorphic
computation cost.

Equation~(\ref{eq:psap}) assigns larger importance scores to filters
that simultaneously exhibit strong predictive contribution, high
activation sensitivity, and limited computational benefit if
removed. Conversely, filters with small importance scores represent
redundant feature maps located in computationally expensive regions
of the network and are therefore selected as pruning candidates.

\subsubsection{Layer-wise Sparsity Allocation}
\label{sec:layer_cap}

Ranking filters according to
Eq.~(\ref{eq:psap}) determines their relative importance within each
layer. However, applying a uniform pruning ratio across all layers
does not account for the non-uniform distribution of activation
sensitivity throughout the network.

To adapt the pruning budget to the characteristics of each layer,
the average activation sensitivity is first computed as

\begin{equation}
\bar{S}_{\ell}
=
\frac{1}{N_{\ell}}
\sum_{i=1}^{N_{\ell}}
S_i,
\label{eq:layer_sensitivity}
\end{equation}

where $N_{\ell}$ denotes the number of filters in layer $\ell$.

The maximum pruning ratio assigned to each layer is then determined
by

\begin{equation}
\mathrm{cap}(\ell)
=
s_0
\left(
1
-
\gamma
\frac
{\bar{S}_{\ell}}
{\displaystyle\max_j\bar{S}_j}
\right),
\label{eq:cap}
\end{equation}

where $s_0$ denotes the baseline sparsity scaling factor (equal to the
target sparsity, e.g., $0.5$ at the $50\%$ operating point) and
$\gamma$ controls the degree of sensitivity-aware protection.
When $\gamma>0$, layers with higher average sensitivity receive
smaller pruning budgets; the resulting per-layer caps may therefore
yield an effective global sparsity below $s_0$.

Layers containing highly sensitive feature representations receive
smaller pruning budgets, whereas layers with lower average
sensitivity are allowed to remove a larger number of filters.
Consequently, the global pruning process becomes adaptive to the
distribution of information throughout the network rather than
applying an identical sparsity ratio to every convolution layer.

After the layer-wise sparsity budgets have been determined, filters
within each layer are ranked according to
Eq.~(\ref{eq:psap}), and those with the smallest importance scores
are removed until the corresponding sparsity constraint defined by
Eq.~(\ref{eq:cap}) is satisfied. The resulting pruned model
$\mathcal{M}_{\mathrm{PSAP}} = \mathcal{T}_{\mathrm{PSAP}}(\mathcal{M}_{\mathrm{Poly}})$
serves as the input to the adaptive polynomial degree
optimization described in the following subsection.

\subsubsection{Fine-Tuning}
\label{sec:finetune}

Structured pruning modifies both the feature representation and the
optimization landscape of the network. Consequently, the pruned
model is fine-tuned to recover the prediction accuracy degraded by
filter removal while preserving the sparsity pattern determined by
the proposed PSAP optimization.

Fine-tuning starts from the pruned model
$\mathcal{M}_{\mathrm{PSAP}}$. The remaining filters retain their
learned parameters, whereas the removed filters are permanently
eliminated from the computation graph. The resulting network is then
fine-tuned using stochastic gradient descent (SGD). This fine-tuning completes
the transformation
$\mathcal{T}_{\mathrm{PSAP}}$
defined in Eq.~(\ref{eq:framework}) and enables the remaining
parameters to compensate for the removed feature representations.

\subsection{Adaptive Polynomial Degree Optimization}
\label{sec:degree}

The pruned network obtained after reliability-aware optimization is
fully compatible with homomorphic encryption (HE) inference.
However, the multiplicative depth required to evaluate the remaining
polynomial activations may still exceed the available CKKS level
budget. The objective of the adaptive polynomial degree
optimization, corresponding to
$\mathcal{T}_{\mathrm{Deg}}$
in Eq.~(\ref{eq:framework}), is therefore to minimize the required
multiplicative depth while preserving the nonlinear behavior of the
most influential activation functions.

\subsubsection{Degree Selection Criterion}
\label{sec:degree_selection}

Following polynomial retraining and reliability-aware pruning, each
activation is represented by $p(x) = ax^2+bx+c$. The contribution of
the quadratic component is quantified through the normalized curvature
ratio

\begin{equation}
\rho
=
\frac{|a|}
{|a|+|b|+|c|},
\label{eq:rho}
\end{equation}

where $\rho\in[0,1]$ represents the relative contribution of the
quadratic term to the overall polynomial.

Small values of $\rho$ indicate that the activation behaves almost
linearly within its learned operating region, whereas larger values
correspond to stronger nonlinear behavior. Because all polynomial
coefficients are jointly trained with the network parameters on the
same data pipeline, the input scales remain consistent across
layers, making the metric stable within a given model despite its
theoretical scale dependence. Consequently, activations
with low curvature ratios can be approximated by linear functions
with limited loss of representational capability.

\subsubsection{Mixed-Degree Assignment}
\label{sec:mixed_degree}

Let
$\mathcal{P} = \{ p_1,p_2,\ldots,p_M \}$
denote the set of polynomial activations in the optimized network.
Each activation is assigned either a first-order or second-order
representation, $d_i \in \{1,2\}$, subject to the available
multiplicative depth constraint

\begin{equation}
\sum_{i=1}^{M}
d_i
+
D_{\mathrm{conv}}
\le
L,
\label{eq:depth_constraint}
\end{equation}

where
$D_{\mathrm{conv}}$
denotes the multiplicative depth consumed by the convolution layers
and $L$ is the available CKKS level budget.

The activations are ranked according to the curvature ratio defined
in Eq.~(\ref{eq:rho}). Degree-two polynomials are preserved for the
activations exhibiting the largest curvature ratios until the depth
constraint is satisfied. The remaining activations are converted
into first-order polynomials $p(x) = bx+c$ using the coefficients
obtained during polynomial retraining. The resulting optimized model
$\mathcal{M}_{\mathrm{Deg}} = \mathcal{T}_{\mathrm{Deg}}(\mathcal{M}_{\mathrm{PSAP}})$
satisfies the multiplicative depth constraints required for
leveled CKKS inference.

\subsubsection{Complexity Analysis}
\label{sec:complexity}

The adaptive degree optimization is performed once during model
preparation and therefore introduces no runtime overhead during
encrypted inference. For $M$ polynomial activations, computing the
curvature ratio in Eq.~(\ref{eq:rho}) is $\mathcal{O}(M)$ and a single
linear traversal assigns the degrees; the sorting step dominates, giving
an overall complexity of $\mathcal{O}(M\log M)$, which is negligible
compared with network training and encrypted inference.

\subsection{Quantization-Aware Training}
\label{sec:qat}

Following adaptive polynomial degree optimization, the network undergoes
a final quantization-aware training (QAT) stage, denoted
$\mathcal{T}_{\mathrm{QAT}}$
in Eq.~(\ref{eq:framework}), which produces the final deployable model
$\mathcal{M}^* = \mathcal{T}_{\mathrm{QAT}}(\mathcal{M}_{\mathrm{Deg}})$.

During this stage, the parameters are fine-tuned under simulated integer
quantization to ensure that the model remains robust against the integer
scaling and rounding artifacts introduced when weights and polynomial
coefficients are encoded into CKKS ciphertexts. Once trained, the final
model $\mathcal{M}^*$ is exported directly into the OpenFHE runtime
environment for leveled homomorphic evaluation.

\subsection{Reliability Modeling}
\label{sec:reliability_framework}

The final optimization stage characterizes the reliability of the
optimized network under transient memory faults. This stage does not
modify the model parameters but provides a systematic framework for
quantifying the effect of the optimization decisions introduced by
$\mathcal{T}_{\mathrm{PSAP}}$
and
$\mathcal{T}_{\mathrm{Deg}}$
on fault propagation, layer vulnerability, and overall inference
robustness.

The reliability modeling framework consists of three components. A
fault model first defines the transient memory faults considered in
this work. A bit-flip injection framework then emulates faults
across different numerical representations, including floating-point
parameters, integer representations, and encrypted CKKS
coefficients. Finally, a set of reliability metrics quantifies the
effect of the injected faults on the optimized model. Together,
these components establish a consistent methodology for comparing
the reliability of different optimization strategies under
identical fault conditions.

The fault model assumes transient hardware faults as independent, uniformly random bit-flips affecting the binary representation of model parameters. The bit error rate (BER) $\beta$ specifies the probability that each bit is flipped independently, producing an expected $P \times b \times \beta$ flipped bits for a model with $P$ parameters of $b$-bit length. The scope focuses on silent data corruptions (SDCs) resulting from single-event upsets in memory holding model parameters and ciphertext coefficients.

To evaluate reliability under different numerical representations, two flip modes are considered. The \emph{int32} flip mode quantizes parameters to 32-bit fixed-point integers before fault injection, where each bit has equal probability of corruption. This mode models CKKS ciphertext coefficient corruption. The \emph{float32} flip mode perturbs parameters in their native IEEE~754 representation, where exponent-bit flips can cause catastrophic magnitude changes and mantissa flips yield bounded variations. For each fault injection campaign, the corrupted model is generated by independently sampling a binary fault mask from a Bernoulli distribution with probability $\beta$ and applying it via an exclusive-or operation.

\subsubsection*{Reliability Metrics}

The optimized models are characterized using complementary metrics that
quantify the impact of transient faults at both the network and layer
levels. The overall reliability of a model is first quantified through the
accuracy degradation under a given bit error rate (BER). Let
$A_{\mathrm{clean}}$ denote the inference accuracy of the fault-free
model and $A_{\mathrm{fault}}(\beta)$ the average accuracy obtained
after fault injection with BER $\beta$. The corresponding accuracy
degradation is defined as

\begin{equation}
\Delta A(\beta)
=
A_{\mathrm{clean}}
-
A_{\mathrm{fault}}(\beta),
\label{eq:acc_drop}
\end{equation}

which measures the global impact of transient faults on encrypted
inference.

To evaluate the contribution of individual layers to the overall
fault behavior, memory faults are independently injected into each
layer while all remaining layers remain fault free. The resulting
layer vulnerability is defined as

\begin{equation}
V_{\ell}
=
A_{\mathrm{clean}}
-
A_{\ell},
\label{eq:layer_vulnerability}
\end{equation}

where $A_{\ell}$ denotes the inference accuracy obtained when only
layer $\ell$ is subjected to fault injection. This metric provides
a quantitative estimate of the contribution of each layer to the
overall fault sensitivity of the network.

Beyond aggregate accuracy, the silent data corruption (SDC) rate
quantifies output errors that occur without any detectable execution
failure. For a test set of $N$ inputs, let $\hat{y}_n$ and
$\hat{y}_n^{\mathrm{clean}}$ denote the predicted labels of input $n$
with and without fault injection. The SDC rate is defined as

\begin{equation}
\mathrm{SDC}(\beta)
=
\frac{1}{N}
\sum_{n=1}^{N}
\mathbb{1}\!\left[
\hat{y}_n \neq \hat{y}_n^{\mathrm{clean}}
\;\wedge\;
\text{execution completes}
\right],
\label{eq:sdc}
\end{equation}

where $\mathbb{1}[\cdot]$ is the indicator function. By excluding
runs that terminate in a detectable execution failure (such as CKKS
ciphertext overflow), the SDC rate isolates the silent mispredictions
that are most dangerous in practice, since they return a confident but
incorrect result with no error signal to the user. Whereas
$\Delta A(\beta)$ measures accuracy loss against the ground truth, the
SDC rate measures divergence from the fault-free prediction.

The analysis additionally considers catastrophic failures, defined as
fault configurations whose accuracy degradation exceeds $10$\,pp, and
the distribution of fault sensitivity across layers
before and after optimization, which reveals whether the framework
merely shifts sensitivity between layers or improves the intrinsic fault
tolerance of the complete network.

\section{Experimental Results}
\label{sec:setup}

\subsection{Experimental Configuration}

Evaluation is performed on ResNet-20 and ResNet-32~\cite{he2016resnet}
across CIFAR-10 and CIFAR-100~\cite{krizhevsky2009cifar}, yielding
four model--dataset configurations. These architectures are the
standard benchmarks in the HE-CNN
literature~\cite{lee2022heresnet,jha2021deepreduce,kim2024hyphen}.

\begin{figure*}[!t]
\centering
\begin{tikzpicture}
\begin{groupplot}[
  group style={group size=2 by 2, horizontal sep=1.4cm, vertical sep=2.0cm},
  width=0.48\textwidth,
  height=4.8cm,
  grid=major,
  grid style={gray!25},
  every axis label/.style={font=\small},
  tick label style={font=\scriptsize},
  every axis plot/.append style={mark size=2.5pt, thick},
  ylabel={Rotation Savings (\%)},
  xtick={20,35,50},
  xmin=15, xmax=55,
]

\nextgroupplot[title={\scriptsize\textbf{(a) ResNet-20 / CIFAR-10}}, ymin=0, ymax=50,
  xlabel={},
  legend style={at={(0.02,0.98)}, anchor=north west, font=\tiny, cells={anchor=west}, draw=gray!40}]
\addplot[color=red!70!black, mark=triangle*, mark options={solid}] coordinates {
  (20, 8.7) (35, 19.8) (50, 26.9)};
\addlegendentry{Magnitude $\alpha{=}0,\gamma{=}0$}
\addplot[color=teal!80!black, mark=square*, mark options={solid}, dashed] coordinates {
  (20, 18.3) (35, 27.6) (50, 40.9)};
\addlegendentry{PSAP $\alpha{=}0.7,\gamma{=}0$}
\addplot[color=blue!70!black, mark=*, mark options={solid}] coordinates {
  (20, 26.4) (35, 26.4) (50, 30.9)};
\addlegendentry{PSAP $\alpha{=}0.7,\gamma{=}0.5$}
\node[font=\tiny, below right, xshift=1pt, yshift=-1pt, color=red!60!black] at (axis cs:50,26.9) {83.8\%};
\node[font=\tiny, above right, xshift=1pt, color=teal!70!black] at (axis cs:50,40.9) {84.2\%};
\node[font=\tiny, above right, xshift=1pt, color=blue!60!black] at (axis cs:50,30.9) {84.2\%};

\nextgroupplot[title={\scriptsize\textbf{(b) ResNet-20 / CIFAR-100}}, ymin=0, ymax=50,
  xlabel={},
  legend style={at={(0.02,0.98)}, anchor=north west, font=\tiny, cells={anchor=west}, draw=gray!40}]
\addplot[color=red!70!black, mark=triangle*, mark options={solid}] coordinates {
  (20, 13.7) (35, 25.3) (50, 30.5)};
\addlegendentry{Magnitude $\alpha{=}0,\gamma{=}0$}
\addplot[color=teal!80!black, mark=square*, mark options={solid}, dashed] coordinates {
  (20, 24.7) (35, 24.4) (50, 39.3)};
\addlegendentry{PSAP $\alpha{=}0.7,\gamma{=}0$}
\addplot[color=blue!70!black, mark=*, mark options={solid}] coordinates {
  (20, 22.4) (35, 33.3) (50, 41.6)};
\addlegendentry{PSAP $\alpha{=}0.7,\gamma{=}0.5$}
\node[font=\tiny, below right, xshift=1pt, yshift=-1pt, color=red!60!black] at (axis cs:50,30.5) {50.5\%};
\node[font=\tiny, below left, xshift=-1pt, yshift=-2pt, color=teal!70!black] at (axis cs:50,39.3) {56.3\%};
\node[font=\tiny, above right, xshift=1pt, yshift=1pt, color=blue!60!black] at (axis cs:50,41.6) {51.5\%};

\nextgroupplot[title={\scriptsize\textbf{(c) ResNet-32 / CIFAR-10}}, ymin=0, ymax=55,
  xlabel={Target Sparsity (\%)},
  legend style={at={(0.02,0.98)}, anchor=north west, font=\tiny, cells={anchor=west}, draw=gray!40}]
\addplot[color=red!70!black, mark=triangle*, mark options={solid}] coordinates {
  (20, 13.0) (35, 21.1) (50, 35.3)};
\addlegendentry{Magnitude $\alpha{=}0,\gamma{=}0$}
\addplot[color=teal!80!black, mark=square*, mark options={solid}, dashed] coordinates {
  (20, 20.7) (35, 26.3) (50, 45.2)};
\addlegendentry{PSAP $\alpha{=}0.7,\gamma{=}0$}
\addplot[color=blue!70!black, mark=*, mark options={solid}] coordinates {
  (20, 20.7) (35, 26.0) (50, 42.7)};
\addlegendentry{PSAP $\alpha{=}0.7,\gamma{=}0.5$}
\node[font=\tiny, below right, xshift=1pt, yshift=-1pt, color=red!60!black] at (axis cs:50,35.3) {88.5\%};
\node[font=\tiny, above right, xshift=1pt, yshift=1pt, color=teal!70!black] at (axis cs:50,45.2) {87.5\%};
\node[font=\tiny, below left, xshift=-1pt, yshift=-4pt, color=blue!60!black] at (axis cs:50,42.7) {87.2\%};

\nextgroupplot[title={\scriptsize\textbf{(d) ResNet-32 / CIFAR-100}}, ymin=0, ymax=55,
  xlabel={Target Sparsity (\%)},
  legend style={at={(0.02,0.98)}, anchor=north west, font=\tiny, cells={anchor=west}, draw=gray!40}]
\addplot[color=red!70!black, mark=triangle*, mark options={solid}] coordinates {
  (20, 18.1) (35, 25.4) (50, 40.3)};
\addlegendentry{Magnitude $\alpha{=}0,\gamma{=}0$}
\addplot[color=teal!80!black, mark=square*, mark options={solid}, dashed] coordinates {
  (20, 26.7) (35, 26.9) (50, 44.9)};
\addlegendentry{PSAP $\alpha{=}0.7,\gamma{=}0$}
\addplot[color=blue!70!black, mark=*, mark options={solid}] coordinates {
  (20, 22.8) (35, 26.9) (50, 42.2)};
\addlegendentry{PSAP $\alpha{=}0.7,\gamma{=}0.5$}
\node[font=\tiny, below right, xshift=2pt, yshift=-6pt, color=red!60!black] at (axis cs:50,40.3) {56.5\%};
\node[font=\tiny, above right, xshift=2pt, yshift=3pt, color=teal!70!black] at (axis cs:50,44.9) {59.0\%};
\node[font=\tiny, left, xshift=-8pt, color=blue!60!black] at (axis cs:50,42.3) {57.9\%};
\end{groupplot}
\end{tikzpicture}
\caption{Rotation savings vs.\ target sparsity for magnitude pruning, PSAP with uniform layer-wise sparsity, and PSAP with sensitivity-protected layer-wise sparsity across all four configurations. Annotated accuracies are reported at the pruning stage (before quantization-aware fine-tuning).}
\label{fig:design_space}
\end{figure*}
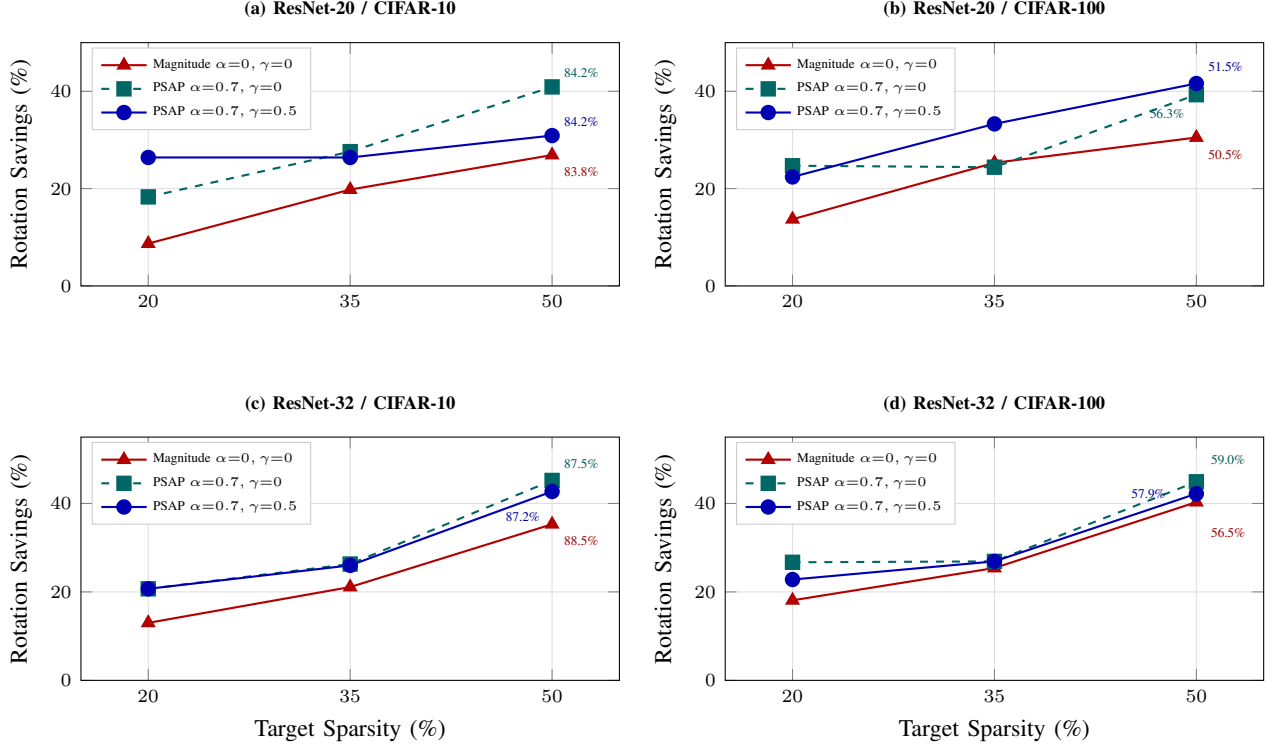

The PSAP hyperparameters were selected empirically and fixed at $\alpha{=}0.7$, $\lambda{=}0.3$,
and $\gamma{=}0$ across all configurations; the $\gamma{=}0.5$ variant
is reported separately as an accuracy-protection alternative
(Fig.~\ref{fig:design_space}). These choices follow simple design
rationales: $\alpha$ is a moderate blending weight that avoids
over-reliance on either raw magnitude or polynomial sensitivity,
$\lambda$ is kept small so that the rotation-cost term modulates
rather than dominates filter ranking, and $\gamma$ controls
sensitivity-aware layer protection ($\gamma{=}0$ disables
sensitivity-aware cap allocation). With $\gamma{=}0$, the per-layer cap in
Eq.~(\ref{eq:cap}) reduces to a uniform budget; consequently, the
reliability gains reported for the default configuration arise from
the filter-importance metric itself, in particular the
activation-sensitivity term $S_i$ and the depth factor
$\Pi_{\ell}$ in Eq.~(\ref{eq:psap}), rather than from
sensitivity-aware cap allocation, which is instead exercised by the
$\gamma{=}0.5$ variant. Baseline models are trained with SGD
(momentum 0.9, Nesterov, weight decay $5{\times}10^{-4}$) for 100
epochs. Phase~1 retrains with Adam for 50 epochs. Phase~2 fine-tunes
with SGD for 50 epochs. Phase~3 retrains with Adam for 50
epochs. Phase~4 applies AdamW with cosine annealing for 80 epochs.
All phases use batch size 128 on a single GPU.

All HE runs use OpenFHE v1.2.0~\cite{albadri2022openfhe} (CKKS,
$N{=}65{,}536$, scale\_bits$~{=}~30$, FIXEDAUTO scaling, 128-bit
security) operating in leveled mode without bootstrapping (single
image per ciphertext). Encrypted inference runs on a shared CPU-only
Linux server (dual AMD EPYC 7352, 503\,GB RAM) with
\texttt{OMP\_NUM\_THREADS=12}.

For the fault injection campaigns, five bit error rates are swept:
$\text{BER} \in \{10^{-7}, 10^{-6}, 10^{-5}, 10^{-4}, 10^{-3}\}$.
Each (configuration, BER) point is evaluated over 15 independent
trials with a freshly sampled fault mask per trial. Results are
reported as the mean over 15 trials with $95\%$ confidence intervals
(Student's $t$, $\mathrm{df}{=}14$), following the statistical fault
injection methodology of~\cite{leveugle2009statistical}.

\subsection{Pipeline Validation}

\subsubsection{Optimization Pipeline}

The proposed optimization framework was first evaluated to verify
that each stage contributes toward producing a computationally
efficient and homomorphic encryption (HE)-compatible model while
maintaining competitive inference accuracy. Table~\ref{tab:per_phase}
summarizes the classification accuracy after each optimization stage
for all evaluated network and dataset configurations.

\begin{table}[!t]
\centering
\caption{Validation of the proposed optimization pipeline. Plaintext
classification accuracy (\%) after each optimization stage at
50\% target sparsity.}
\label{tab:per_phase}
\renewcommand{\arraystretch}{1.10}
\setlength{\tabcolsep}{3pt}
\scriptsize
\begin{tabular}{@{}p{3.6cm}cccc@{}}
\toprule
\textbf{Phase} & \textbf{R20/C10} & \textbf{R20/C100} & \textbf{R32/C10} & \textbf{R32/C100} \\
\midrule
Baseline (ReLU)                  & 91.34 & 68.73 & 93.44 & 70.09 \\
Polynomial Activation Modeling   & 86.26 & 65.44 & 92.06 & 67.39 \\
PSAP                             & 84.17 & 56.28 & 87.52 & 58.95 \\
Adaptive Degree Optimization     & 83.21 & 55.14 & 86.23 & 57.01 \\
Quantization-Aware Training      & 87.31 & 59.68 & 89.58 & 61.43 \\
\midrule
Encrypted CKKS Inference         & 87.31 & 59.68 & 89.58 & 61.43 \\
\bottomrule
\end{tabular}
\end{table}

Replacing ReLU with trainable degree-two polynomial activations
introduces the largest single-stage degradation for the CIFAR-10
models (91.34\% to 86.26\% for ResNet-20), reflecting the approximation
error of satisfying the multiplicative constraints of the
Cheon--Kim--Kim--Song (CKKS) scheme; for the CIFAR-100 configurations
the polynomial transition is milder and the dominant drop instead occurs
during PSAP pruning. Because PSAP removes structurally redundant filters
while preserving those with high activation sensitivity, the degradation
after 50\% pruning remains controlled, and adaptive mixed-degree
optimization adds less than two percentage points of further loss while
substantially reducing multiplicative depth.

The final quantization-aware training stage recovers 3.35--4.54
percentage points across configurations, yielding plaintext accuracies
of 87.31\%, 59.68\%, 89.58\%, and 61.43\%. These match the encrypted
CKKS accuracies exactly, confirming that the adopted parameterization
and quantization introduce no additional prediction error beyond that
modeled during training. The four-stage pipeline thus transforms a
conventional network into an HE-compatible model while preserving
competitive accuracy, validating the framework of
Section~\ref{sec:method}.

\subsubsection{Computational Efficiency}
\label{sec:efficiency}

The computational efficiency of the proposed optimization framework
was evaluated by quantifying its impact on the computational and
memory requirements of homomorphic encrypted inference.
Table~\ref{tab:pipeline} summarizes the resource consumption before
and after applying the complete optimization pipeline, while
Fig.~\ref{fig:design_space} illustrates the relationship between
rotation savings, target sparsity, and prediction accuracy across
different pruning strategies.

\begin{table}[!t]
\centering
\caption{Computational efficiency of the proposed optimization
framework. Results compare the polynomial model before structured
pruning (Phase~1) with the final PSAP-optimized model.}
\label{tab:pipeline}
\renewcommand{\arraystretch}{1.10}
\setlength{\tabcolsep}{3pt}
\scriptsize
\begin{tabular}{llccc}
\toprule
\textbf{Model} & \textbf{Metric} & \textbf{Unpr. (Ph.1)} & \textbf{PSAP} & \textbf{$\Delta$} \\
\midrule

\multirow{5}{*}{\rotatebox{90}{\textbf{R-20}}}
  & Rotations (M) & 40.8 & \textbf{24.1} & $-$40.9\% \\
  & Latency (s/img) & 1{,}460 & \textbf{1{,}181} & $-$19.1\% \\
  & Peak Mem. (GB) & 43.7 & \textbf{40.0} & $-$8.5\% \\
  & HE Context (GB) & 32.4 & \textbf{30.1} & $-$7.1\% \\
  & CT Size (MB) & 46.0 & \textbf{44.0} & $-$4.3\% \\
\midrule

\multirow{5}{*}{\rotatebox{90}{\textbf{R-32}}}
  & Rotations (M) & 69.1 & \textbf{37.9} & $-$45.2\% \\
  & Latency (s/img) & \multirow{4}{*}{\textit{Infeasible}} & 2{,}244 & --- \\
  & Peak Mem. (GB) & & 25.7 & --- \\
  & HE Context (GB) & & 19.3 & --- \\
  & CT Size (MB) & & 58.0 & --- \\
\bottomrule
\end{tabular}
\end{table}

The proposed framework substantially reduces the cost of encrypted
inference across both architectures. For ResNet-20, Halevi--Shoup
rotations drop by 40.9\% (40.8M to 24.1M), lowering estimated latency
from 1460\,s to 1181\,s per image and reducing peak memory, HE context,
and ciphertext sizes by 8.5\%, 7.1\%, and 4.3\%, respectively. The
benefits are more pronounced for ResNet-32, where rotations fall by
45.2\% (69.1M to 37.9M) and, more importantly, the multiplicative depth
decreases from 66 to 56 levels. This reduces the required ciphertext
modulus from 2040 to 1740 bits, fitting within the 1760-bit limit of a
ring dimension of 65\,536 at 128-bit security and enabling
bootstrapping-free encrypted inference; the original polynomial network
exceeds the modulus chain and cannot execute under the same
configuration.

The absolute memory, context, and ciphertext-size values in
Table~\ref{tab:pipeline} depend on the modulus-chain length, the rotation
key set, and the packing layout, and are therefore not monotonic in depth
(the optimized ResNet-32 reports lower peak memory and context size than
ResNet-20 despite a larger per-ciphertext size); the per-architecture
relative reductions provide the meaningful comparison.

Figure~\ref{fig:design_space} compares PSAP against conventional
magnitude pruning over target sparsities from 20\% to 50\%. PSAP
consistently achieves larger rotation reductions at comparable or higher
accuracy: at 50\% sparsity, rotation savings reach 40.9\% and 45.2\% for
ResNet-20/CIFAR-10 and ResNet-32/CIFAR-10, versus 26.9\% and 35.3\% for
magnitude pruning, with similar gains on CIFAR-100 (30.5\% to 39.3\% for
ResNet-20, 40.3\% to 44.9\% for ResNet-32). Sensitivity-protected
layer-wise allocation ($\gamma=0.5$) offers an alternative operating
point that trades slightly lower rotation savings for higher accuracy at
aggressive sparsity, demonstrating PSAP's flexibility in balancing
efficiency and predictive performance.

\subsection{Reliability Characterization}

\subsubsection{Global Fault Tolerance}

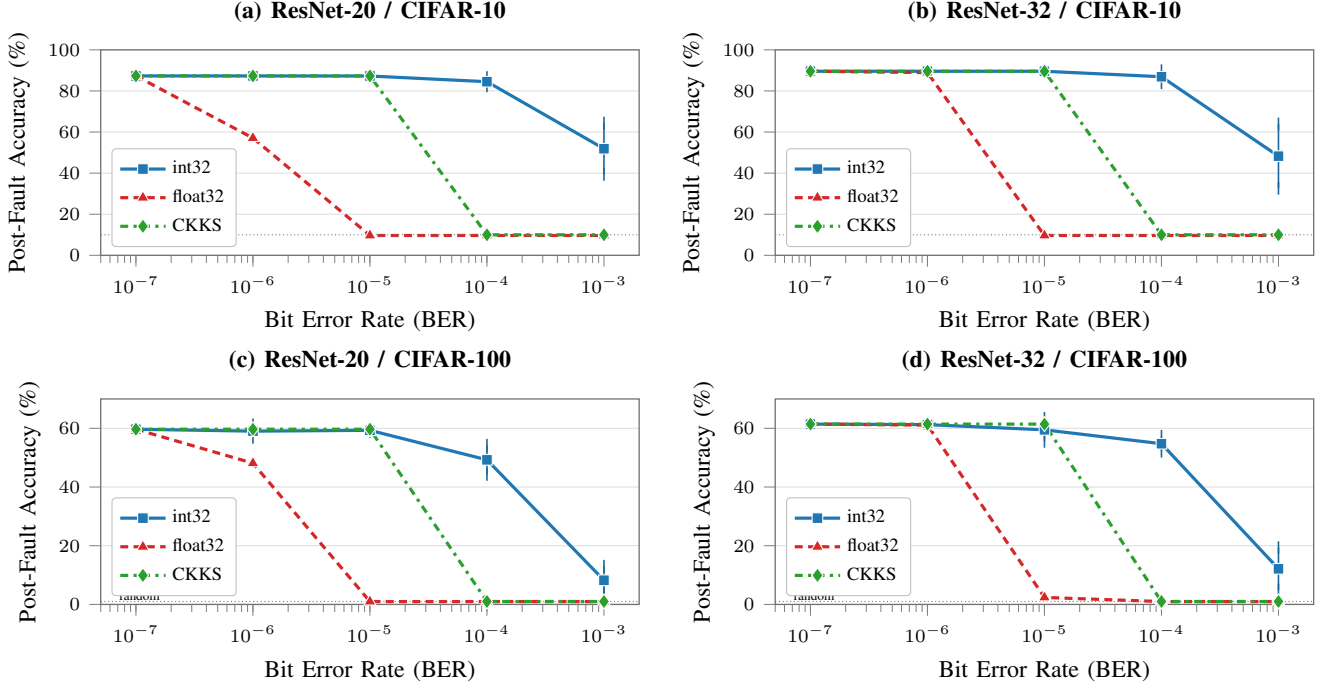
\begin{figure*}[!t]
\centering
\begin{tikzpicture}
\begin{groupplot}[
    faultbase,
    group style={
        group size=2 by 2,
        horizontal sep=1.8cm,
        vertical sep=1.9cm,
    },
    ylabel={Post-Fault Accuracy (\%)},
]

\nextgroupplot[
    title={(a) ResNet-20 / CIFAR-10},
    ymin=0, ymax=100,
    legend style={at={(0.02,0.04)}, anchor=south west},
]

\addplot[int32line, ebcfg] coordinates {
    (1e-7, 87.31) +- (0,0)
    (1e-6, 87.31) +- (0,0.07)
    (1e-5, 87.27) +- (0,0.24)
    (1e-4, 84.50) +- (0,2.01)
    (1e-3, 51.88) +- (0,12.47)
};
\addlegendentry{int32}

\addplot[float32line] coordinates {
    (1e-7, 87.31) (1e-6, 57.08) (1e-5, 9.67) (1e-4, 9.67) (1e-3, 9.67)
};
\addlegendentry{float32}

\addplot[randline] coordinates {(5e-8, 10) (2e-3, 10)};
\node[font=\tiny, anchor=west] at (axis cs:6e-8, 12) {random};

\addplot[ckksline] coordinates {
    (1e-7, 87.31) (1e-6, 87.31) (1e-5, 87.31) (1e-4, 10) (1e-3, 10)
};
\addlegendentry{CKKS}

\nextgroupplot[
    title={(b) ResNet-32 / CIFAR-10},
    ymin=0, ymax=100,
    legend style={at={(0.02,0.04)}, anchor=south west},
]

\addplot[int32line, ebcfg] coordinates {
    (1e-7, 89.58) +- (0,0)
    (1e-6, 89.58) +- (0,0)
    (1e-5, 89.58) +- (0,0)
    (1e-4, 86.91) +- (0,2.92)
    (1e-3, 48.26) +- (0,15.64)
};
\addlegendentry{int32}

\addplot[float32line] coordinates {
    (1e-7, 89.58) (1e-6, 88.91) (1e-5, 9.67) (1e-4, 9.67) (1e-3, 9.67)
};
\addlegendentry{float32}
\addplot[randline] coordinates {(5e-8, 10) (2e-3, 10)};
\node[font=\tiny, anchor=west] at (axis cs:6e-8, 12) {random};

\addplot[ckksline] coordinates {
    (1e-7, 89.58) (1e-6, 89.58) (1e-5, 89.58) (1e-4, 10) (1e-3, 10)
};
\addlegendentry{CKKS}

\nextgroupplot[
    title={(c) ResNet-20 / CIFAR-100},
    ymin=0, ymax=70,
    legend style={at={(0.02,0.04)}, anchor=south west},
]

\addplot[int32line, ebcfg] coordinates {
    (1e-7, 59.68) +- (0,0)
    (1e-6, 59.03) +- (0,2.11)
    (1e-5, 59.31) +- (0,0.36)
    (1e-4, 49.26) +- (0,4.95)
    (1e-3, 8.25) +- (0,4.75)
};
\addlegendentry{int32}

\addplot[float32line] coordinates {
    (1e-7, 59.68) (1e-6, 48.11) (1e-5, 0.98) (1e-4, 0.98) (1e-3, 0.98)
};
\addlegendentry{float32}
\addplot[randline] coordinates {(5e-8, 1) (2e-3, 1)};
\node[font=\tiny, anchor=west] at (axis cs:6e-8, 3) {random};

\addplot[ckksline] coordinates {
    (1e-7, 59.68) (1e-6, 59.68) (1e-5, 59.68) (1e-4, 1) (1e-3, 1)
};
\addlegendentry{CKKS}

\nextgroupplot[
    title={(d) ResNet-32 / CIFAR-100},
    ymin=0, ymax=70,
    legend style={at={(0.02,0.04)}, anchor=south west},
]

\addplot[int32line, ebcfg] coordinates {
    (1e-7, 61.43) +- (0,0)
    (1e-6, 61.25) +- (0,0.20)
    (1e-5, 59.48) +- (0,3.93)
    (1e-4, 54.75) +- (0,2.53)
    (1e-3, 12.13) +- (0,7.21)
};
\addlegendentry{int32}

\addplot[float32line] coordinates {
    (1e-7, 61.43) (1e-6, 61.09) (1e-5, 2.38) (1e-4, 0.98) (1e-3, 0.98)
};
\addlegendentry{float32}
\addplot[randline] coordinates {(5e-8, 1) (2e-3, 1)};
\node[font=\tiny, anchor=west] at (axis cs:6e-8, 3) {random};

\addplot[ckksline] coordinates {
    (1e-7, 61.43) (1e-6, 61.43) (1e-5, 61.43) (1e-4, 1) (1e-3, 1)
};
\addlegendentry{CKKS}

\end{groupplot}
\end{tikzpicture}
\caption{Post-fault accuracy vs.\ BER for int32, float32, and CKKS bit-flip modes across all four configurations. Error bars on the int32 curves are 95\% CIs over 15 trials (Section~\ref{sec:setup}); float32 and CKKS are deterministic.}
\label{fig:ber_accuracy_all}
\end{figure*}

Figure~\ref{fig:ber_accuracy_all} presents the classification accuracy
under increasing bit error rates (BERs) for all evaluated
model--dataset configurations. The results characterize the global
fault tolerance of the optimized models and quantify how transient
memory faults affect inference reliability under different network
depths and classification complexities.

Across all configurations, the models remain highly resilient at low
fault rates. For BERs up to $10^{-6}$, the classification accuracy
remains close to the fault-free baseline, indicating that isolated
memory bit flips are largely absorbed by the inherent redundancy of
the network. A noticeable degradation begins at BER~=~$10^{-5}$, and
the most pronounced loss occurs between $10^{-4}$ and $10^{-3}$, where
the probability of simultaneously perturbing multiple critical weights
becomes high enough to disrupt the learned representations.

The rate of degradation depends on both architecture and task. CIFAR-10
models consistently retain higher accuracy than their CIFAR-100
counterparts, as the larger output space narrows the classification
margin and increases sensitivity to perturbations. A similar dependency
holds for depth: the deeper ResNet-32 preserves accuracy over a wider
BER range than ResNet-20 at moderate fault rates, since its additional
residual blocks distribute isolated perturbations across more feature
transformations. Once the BER reaches $10^{-3}$, this redundancy is no
longer sufficient and both architectures degrade substantially. Overall,
the optimized models degrade gradually rather than abruptly under
realistic transient fault rates, establishing the global
fault-tolerance baseline used for the layer-wise analyses that follow.

\subsubsection{Silent Data Corruption and Logit Stability}

While classification accuracy quantifies the final impact of memory
faults, it does not capture the evolution of internal numerical errors
before misclassification occurs. Silent Data Corruption (SDC) and logit
Mean Absolute Error (MAE) are therefore analyzed to characterize fault
propagation from parameter perturbation to output corruption.

\begin{figure*}[!t]
\centering
\begin{tikzpicture}
\begin{groupplot}[
    faultbase,
    group style={
        group size=2 by 2,
        horizontal sep=1.8cm,
        vertical sep=1.9cm,
    },
    ylabel={SDC Rate (\%)},
    ymin=0, ymax=112,
]

\nextgroupplot[
    title={(a) ResNet-20 / CIFAR-10},
    legend style={at={(0.02,0.98)}, anchor=north west},
]
\addplot[int32line] coordinates {
    (1e-7, 0.0) (1e-6, 0.07) (1e-5, 0.60) (1e-4, 4.73) (1e-3, 37.99)
};
\addlegendentry{int32}

\addplot[ckksline] coordinates {
    (1e-7, 4.0) (1e-6, 4.0) (1e-5, 4.0) (1e-4, 100) (1e-3, 100)
};
\addlegendentry{CKKS}
\addplot[float32line] coordinates {
    (1e-7, 0.0) (1e-6, 30.77) (1e-5, 77.49) (1e-4, 77.49) (1e-3, 77.49)
};
\addlegendentry{float32}

\nextgroupplot[
    title={(b) ResNet-32 / CIFAR-10},
    legend style={at={(0.02,0.98)}, anchor=north west},
]
\addplot[int32line] coordinates {
    (1e-7, 0.0) (1e-6, 0.0) (1e-5, 0.0) (1e-4, 4.47) (1e-3, 43.57)
};
\addlegendentry{int32}

\addplot[ckksline] coordinates {
    (1e-7, 0.0) (1e-6, 0.0) (1e-5, 0.0) (1e-4, 100) (1e-3, 100)
};
\addlegendentry{CKKS}
\addplot[float32line] coordinates {
    (1e-7, 0.0) (1e-6, 0.17) (1e-5, 78.47) (1e-4, 78.47) (1e-3, 78.47)
};
\addlegendentry{float32}

\nextgroupplot[
    title={(c) ResNet-20 / CIFAR-100},
    legend style={at={(0.02,0.98)}, anchor=north west},
]
\addplot[int32line] coordinates {
    (1e-7, 0.0) (1e-6, 1.12) (1e-5, 1.90) (1e-4, 14.51) (1e-3, 52.70)
};
\addlegendentry{int32}

\addplot[ckksline] coordinates {
    (1e-7, 6.0) (1e-6, 6.0) (1e-5, 6.0) (1e-4, 100) (1e-3, 100)
};
\addlegendentry{CKKS}
\addplot[float32line] coordinates {
    (1e-7, 0.0) (1e-6, 15.89) (1e-5, 57.76) (1e-4, 57.76) (1e-3, 57.76)
};
\addlegendentry{float32}

\nextgroupplot[
    title={(d) ResNet-32 / CIFAR-100},
    legend style={at={(0.02,0.98)}, anchor=north west},
]
\addplot[int32line] coordinates {
    (1e-7, 0.0) (1e-6, 0.51) (1e-5, 3.31) (1e-4, 10.32) (1e-3, 50.93)
};
\addlegendentry{int32}

\addplot[ckksline] coordinates {
    (1e-7, 10.0) (1e-6, 10.0) (1e-5, 13.33) (1e-4, 100) (1e-3, 100)
};
\addlegendentry{CKKS}
\addplot[float32line] coordinates {
    (1e-7, 0.0) (1e-6, 0.61) (1e-5, 58.49) (1e-4, 59.86) (1e-3, 59.86)
};
\addlegendentry{float32}

\end{groupplot}
\end{tikzpicture}
\caption{Silent Data Corruption (SDC) rate vs.\ BER for int32, float32, and CKKS modes across all four configurations.}
\label{fig:sdc_rate_all}
\end{figure*}
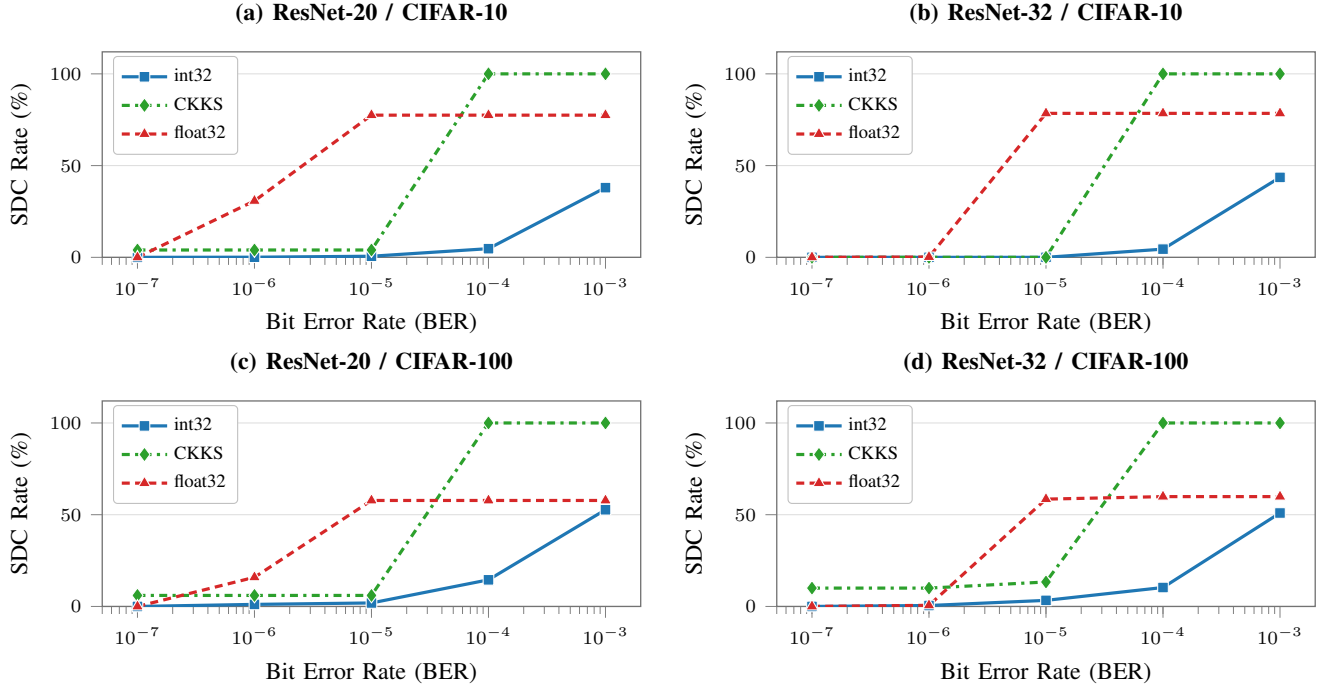

Figure~\ref{fig:sdc_rate_all} presents the SDC rate
[Eq.~(\ref{eq:sdc})] versus BER for the
three numerical representations. The int32 implementation exhibits a
gradual increase in silent failures, closely following the accuracy
trends above. At BER~=~$10^{-5}$, the SDC rate stays below 2\% for the
ResNet-20 models and reaches 3.31\% for ResNet-32/CIFAR-100, rising to
4.73\% and 4.47\% for the CIFAR-10 models and 14.51\% and 10.32\% for
the CIFAR-100 models at BER~=~$10^{-4}$, reflecting the smaller
classification margins of the more challenging dataset.

The most significant increase occurs at BER~=~$10^{-3}$, where SDC
rates reach 37.99\% and 43.57\% for ResNet-20 and ResNet-32 on
CIFAR-10 and exceed 50\% for both CIFAR-100 models. The majority of
failures at high BER therefore correspond to silent prediction errors
rather than detectable execution failures, emphasizing the importance
of evaluating reliability beyond classification accuracy alone.

The float32 representation behaves differently: a single exponent-bit
corruption generates extremely large values that propagate rapidly, so
the SDC rate rises sharply at BER~=~$10^{-6}$ and saturates by
$10^{-5}$ (77.49\% and 78.47\% for CIFAR-10; 57.76\% and 58.49\% for
CIFAR-100) as the network collapses to random prediction.

The CKKS implementation shows a binary reliability profile. No
measurable SDC increase is observed up to BER~=~$10^{-5}$, confirming
that the ciphertext noise budget absorbs low-rate perturbations; the
non-zero CKKS baseline at the lowest BER (e.g., 4\% for R20/C10) reflects
the numerical gap between the quantized plaintext model and CKKS
execution rather than injected faults. During decryption the scaling
factor $\Delta$ separates message bits from low-order noise, acting as a
built-in error filter as long as the perturbation stays within the noise
budget. Once the accumulated error exceeds the modulus at
BER~=~$10^{-4}$, all encrypted executions terminate due to ciphertext
overflow, producing a 100\% failure rate rather than silent corruption.

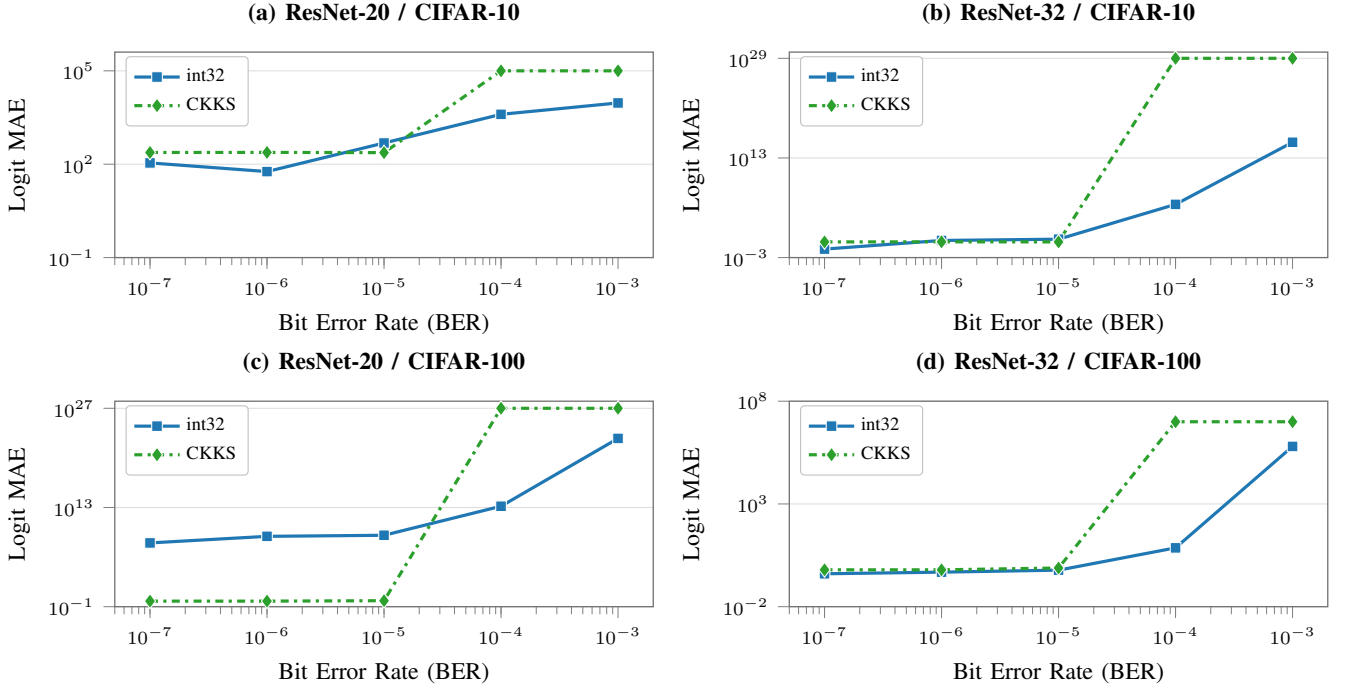
\begin{figure*}[!t]
\centering
\begin{tikzpicture}
\begin{groupplot}[
    faultbase,
    ymode=log,
    group style={
        group size=2 by 2,
        horizontal sep=1.8cm,
        vertical sep=1.9cm,
    },
    ylabel={Logit MAE},
]

\nextgroupplot[
    title={(a) ResNet-20 / CIFAR-10},
    ymin=1e-1, ymax=4e5,
    legend style={at={(0.02,0.98)}, anchor=north west},
]
\addplot[int32line] coordinates {
    (1e-7, 1.1e2) (1e-6, 5.8e1) (1e-5, 4.8e2) (1e-4, 4.0e3) (1e-3, 9.3e3)
};
\addlegendentry{int32}

\addplot[ckksline] coordinates {
    (1e-7, 238.1) (1e-6, 239.6) (1e-5, 235.5) (1e-4, 1e5) (1e-3, 1e5)
};
\addlegendentry{CKKS}

\nextgroupplot[
    title={(b) ResNet-32 / CIFAR-10},
    ymin=1e-3, ymax=1e30,
    legend style={at={(0.02,0.98)}, anchor=north west},
]
\addplot[int32line] coordinates {
    (1e-7, 0.023) (1e-6, 0.564) (1e-5, 0.900) (1e-4, 3.5e5) (1e-3, 3.3e15)
};
\addlegendentry{int32}

\addplot[ckksline] coordinates {
    (1e-7, 0.328) (1e-6, 0.340) (1e-5, 0.328) (1e-4, 1e29) (1e-3, 1e29)
};
\addlegendentry{CKKS}

\nextgroupplot[
    title={(c) ResNet-20 / CIFAR-100},
    ymin=1e-1, ymax=1e28,
    legend style={at={(0.02,0.98)}, anchor=north west},
]
\addplot[int32line] coordinates {
    (1e-7, 1.0e8) (1e-6, 8.5e8) (1e-5, 1.2e9) (1e-4, 1.5e13) (1e-3, 5.6e22)
};
\addlegendentry{int32}

\addplot[ckksline] coordinates {
    (1e-7, 0.61) (1e-6, 0.61) (1e-5, 0.70) (1e-4, 1e27) (1e-3, 1e27)
};
\addlegendentry{CKKS}

\nextgroupplot[
    title={(d) ResNet-32 / CIFAR-100},
    ymin=1e-2, ymax=1e8,
    legend style={at={(0.02,0.98)}, anchor=north west},
]
\addplot[int32line] coordinates {
    (1e-7, 4.0e-1) (1e-6, 4.8e-1) (1e-5, 6.0e-1) (1e-4, 7.3) (1e-3, 6.3e5)
};
\addlegendentry{int32}

\addplot[ckksline] coordinates {
    (1e-7, 0.623) (1e-6, 0.623) (1e-5, 0.764) (1e-4, 1e7) (1e-3, 1e7)
};
\addlegendentry{CKKS}

\end{groupplot}
\end{tikzpicture}
\caption{Logit MAE (log scale) vs.\ BER for int32 and CKKS modes across all four configurations.}
\label{fig:logit_mae_all}
\end{figure*}

The logit MAE in Fig.~\ref{fig:logit_mae_all} provides additional
insight, measuring the numerical deviation of the outputs before the
final classification stage. The logit error generally increases with
BER, demonstrating continuous accumulation of perturbations through
forward propagation; the minor non-monotonicity at the lowest fault
rates reflects the variance of the trial-averaged baseline.

The magnitude of the logit perturbation spans more than twenty
orders of magnitude across the evaluated BER range. For
ResNet-32/CIFAR-10, the logit MAE increases from approximately
$2.3\times10^{-2}$ at BER~=~$10^{-7}$ to
$3.3\times10^{15}$ at BER~=~$10^{-3}$. An even larger increase is
observed for ResNet-20/CIFAR-100, where the MAE grows from
approximately $10^{8}$ to $5.6\times10^{22}$. Similar trends are
observed for the remaining configurations, indicating that internal
numerical errors accumulate much faster than the corresponding
degradation observed in prediction accuracy. The absolute logit-MAE
values are reported in raw logit units and are therefore
model-dependent; in particular, the near-fault-free baselines differ
by several orders of magnitude across configurations because they are
dominated by the rare high-order-bit flips captured within the
averaging across trials. Baseline magnitudes should accordingly be compared
within a panel rather than across panels.

Float32 results are intentionally omitted from the logit analysis
because exponent-bit corruption frequently produces NaN or Inf
values, making the MAE undefined. In contrast, the CKKS
implementation maintains a nearly constant baseline error throughout
the safe operating region, with MAE values remaining approximately
at 238 for ResNet-20/CIFAR-10, 0.33 for ResNet-32/CIFAR-10, and between 0.6 and 0.7
for the CIFAR-100 models. These baselines are reported in raw logit units, so their absolute magnitude is model-dependent; what matters is that each remains flat throughout the safe region. Once BER reaches $10^{-4}$, ciphertext
overflow causes an abrupt increase in the measured error, consistent
with the execution failures observed during encrypted inference.

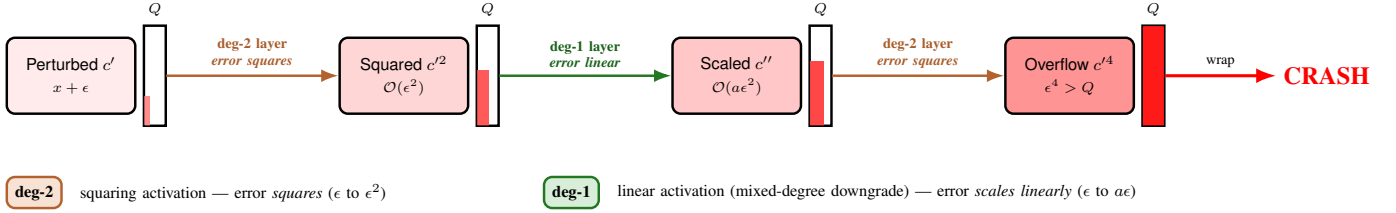
\begin{figure*}[!t]
\centering
\resizebox{\textwidth}{!}{%
\begin{tikzpicture}[
    ct/.style={draw, very thick, rectangle, rounded corners, minimum height=1.3cm,
        minimum width=2.2cm, align=center, font=\footnotesize\sffamily, inner sep=5pt},
    therm/.style={draw, thick, rectangle, minimum width=0.36cm, minimum height=1.7cm},
    deg2arrow/.style={line width=1.2pt, -Latex, draw=cdeg2!80!black},
    deg1arrow/.style={line width=1.2pt, -Latex, draw=cdeg1!75!black},
]

\node[ct, fill=red!8] (c0) at (0,0) {Perturbed $c'$\\[2pt]\scriptsize $x+\epsilon$};
\node[therm, right=0.12cm of c0, anchor=west, yshift=0] (t0) {};
\fill[red!45] (t0.south west) rectangle ($(t0.south west)!0.30!(t0.north east)$);
\draw[thick] (t0.south west) rectangle (t0.north east);
\node[font=\scriptsize\bfseries, above=1pt of t0.north] {$Q$};

\node[ct, fill=red!16] (c1) at (5.7,0) {Squared $c'^2$\\[2pt]\scriptsize $\mathcal{O}(\epsilon^2)$};
\node[therm, right=0.12cm of c1, anchor=west, yshift=0] (t1) {};
\fill[red!65] (t1.south west) rectangle ($(t1.south west)!0.55!(t1.north east)$);
\draw[thick] (t1.south west) rectangle (t1.north east);
\node[font=\scriptsize\bfseries, above=1pt of t1.north] {$Q$};
\draw[deg2arrow] (t0.east) -- node[above, font=\scriptsize\bfseries, text=cdeg2!75!black, align=center]
    {deg-2 layer\\\itshape error squares} (c1.west);

\node[ct, fill=red!22] (c2) at (11.4,0) {Scaled $c''$\\[2pt]\scriptsize $\mathcal{O}(a\epsilon^2)$};
\node[therm, right=0.12cm of c2, anchor=west, yshift=0] (t2) {};
\fill[red!72] (t2.south west) rectangle ($(t2.south west)!0.64!(t2.north east)$);
\draw[thick] (t2.south west) rectangle (t2.north east);
\node[font=\scriptsize\bfseries, above=1pt of t2.north] {$Q$};
\draw[deg1arrow] (t1.east) -- node[above, font=\scriptsize\bfseries, text=cdeg1!60!black, align=center]
    {deg-1 layer\\\itshape error linear} (c2.west);

\node[ct, fill=red!42] (c3) at (17.1,0) {Overflow $c'^4$\\[2pt]\scriptsize $\epsilon^4 > Q$};
\node[therm, right=0.12cm of c3, anchor=west, yshift=0] (t3) {};
\fill[red!90] (t3.south west) rectangle (t3.north east);
\draw[thick] (t3.south west) rectangle (t3.north east);
\node[font=\scriptsize\bfseries, above=1pt of t3.north] {$Q$};
\draw[deg2arrow] (t2.east) -- node[above, font=\scriptsize\bfseries, text=cdeg2!75!black, align=center]
    {deg-2 layer\\\itshape error squares} (c3.west);
\node[font=\large\bfseries, text=red] (crash) at (21.5,0) {CRASH};
\draw[line width=1.5pt, -Latex, red] (t3.east) -- node[above, font=\scriptsize, text=black] {wrap} (crash.west);

\node[draw, very thick, rounded corners, fill=cdeg2!22, draw=cdeg2!80!black,
        font=\footnotesize\bfseries, inner sep=4pt, anchor=west] (lg2) at ($(c0.west)+(0,-2.0)$) {deg-2};
\node[right=5pt of lg2, font=\footnotesize, anchor=west]
    {squaring activation --- error \emph{squares} ($\epsilon$ to $\epsilon^2$)};
\node[draw, very thick, rounded corners, fill=cdeg1!18, draw=cdeg1!75!black,
    font=\footnotesize\bfseries, inner sep=4pt, anchor=west] (lg1) at ($(lg2.west)+(9.2,0)$) {deg-1};
\node[right=5pt of lg1, font=\footnotesize, anchor=west]
    {linear activation (mixed-degree downgrade) --- error \emph{scales linearly} ($\epsilon$ to $a\epsilon$)};
\end{tikzpicture}%
}
\caption{The Overflow Cascade mechanism: scale-bit growth under mixed-degree activation allocation in CKKS. Forward inference: sequential CKKS activations compound the bit-flip error until it exceeds the modulus $Q$.}
\label{fig:overflow_cascade}
\end{figure*}

Figure~\ref{fig:overflow_cascade} illustrates the underlying
mechanism responsible for this behavior. Small perturbations
introduced into ciphertext coefficients are repeatedly transformed
through successive polynomial activations. Degree-two activations
square the propagated error, whereas degree-one activations introduced
by the adaptive mixed-degree optimization are expected to increase the
error only linearly. This mechanism suggests that, beyond reducing
multiplicative depth, the mixed-degree strategy should also slow the
accumulation of ciphertext noise and thereby delay the point at which
the available modulus is exhausted. Once the accumulated error exceeds
the ciphertext modulus, modular wrap-around corrupts the encrypted
computation, leading to deterministic execution failure. The schematic
in Fig.~\ref{fig:overflow_cascade} is illustrative rather than a
fitted model; nevertheless, the observed CKKS behavior, namely a flat
error within the safe region followed by an abrupt failure once the
modulus is exhausted, is consistent with this error-propagation
mechanism.

\subsubsection{Architecture-Level Reliability}

The impact of transient faults depends not only on the bit error rate but
also on the network architecture and task complexity. Under int32 fault
injection, the additional depth of ResNet-32 does not increase its
sensitivity: its degradation at BER~=~$10^{-4}$ matches ResNet-20 on
CIFAR-10 (about 2.7 versus 2.8 percentage points), as the extra parameters
are offset by the representational redundancy of its residual structure.
The role of depth becomes clearer under float32
perturbations, where ResNet-32 retains 88.91\% accuracy at
BER~=~$10^{-6}$ versus 57.08\% for ResNet-20. Network depth alone
therefore does not determine reliability; it is the interaction between
architectural redundancy, numerical representation, and task complexity
that governs the fault response of encrypted neural networks.

\subsubsection{Error Propagation Analysis}

This subsection relates the external fault manifestations to the
underlying propagation mechanism. At low BERs,
individual bit flips produce small, localized perturbations that are
largely attenuated by network redundancy. As the BER increases,
high-order bit corruptions produce larger perturbations that propagate
and amplify through successive convolutional layers and polynomial
activations, corresponding to the rapid rise in SDC rate and logit MAE
between BER~=~$10^{-5}$ and $10^{-4}$, which marks the practical
reliability boundary.

The mechanism differs by representation. Under int32 arithmetic,
accumulated errors remain bounded, producing gradual accuracy loss.
Float32 instead degrades abruptly, as exponent-bit corruption generates
extremely large values leading to NaN or Inf activations. In CKKS, small
perturbations are initially absorbed by the ciphertext noise budget
(no measurable degradation up to BER~=~$10^{-5}$); as illustrated in
Fig.~\ref{fig:overflow_cascade}, successive homomorphic multiplications
accumulate error until the modulus is exhausted, after which modular
wrap-around causes deterministic execution failure. The agreement between
plaintext injection, direct CKKS experiments, and the overflow behavior
confirms that the framework captures the progression from localized bit
corruption to system-level failure.

\subsection{Layer-wise Reliability}

\subsubsection{Layer Criticality}

The global fault analysis characterizes overall robustness but does not
reveal which components dominate the degradation. To identify the
structural origin of fault propagation, transient bit flips are injected
into each convolutional layer independently while all others remain fault
free, using the int32 representation at BERs of $10^{-6}$, $10^{-5}$, and
$10^{-4}$ for all four configurations.

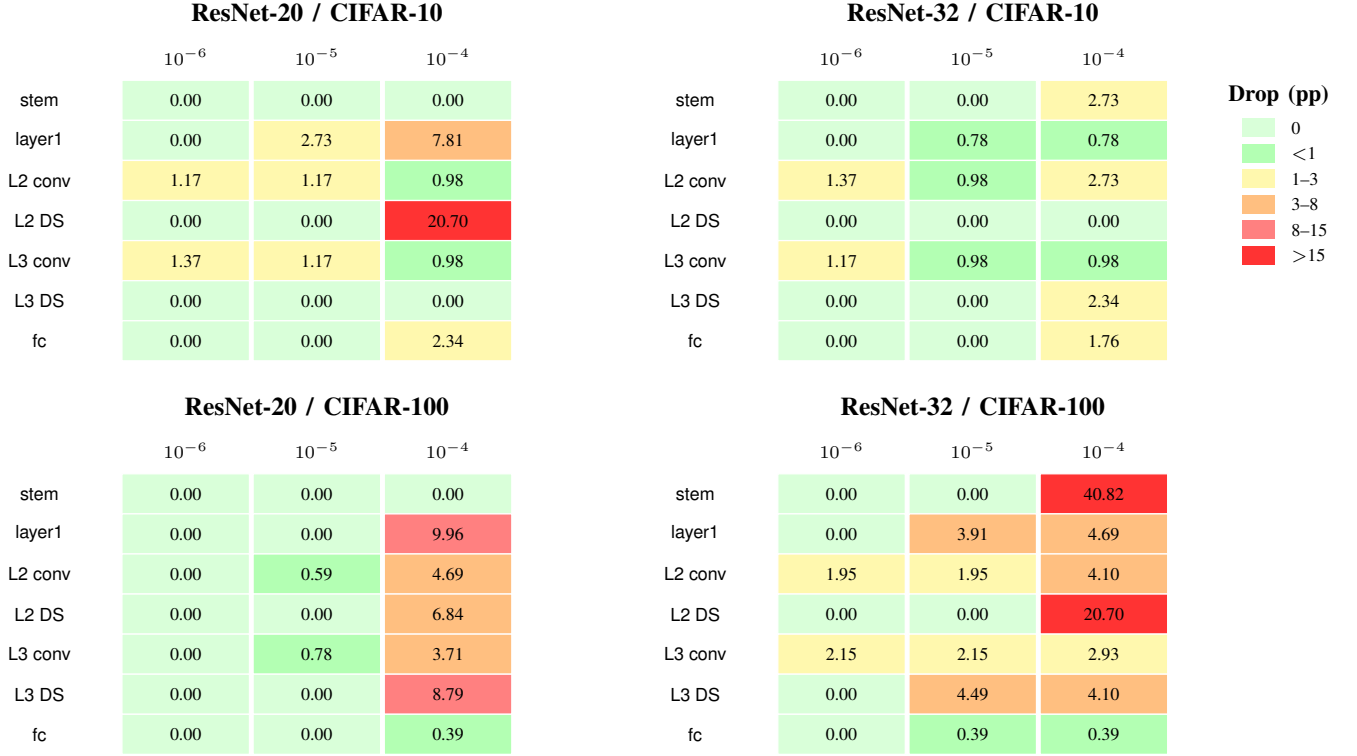
\begin{figure*}[!t]
\centering
\resizebox{\textwidth}{!}{%
\begin{tikzpicture}[
    cell/.style={minimum width=1.75cm, minimum height=0.55cm, align=center, font=\scriptsize, inner sep=1pt, draw=white, line width=0.5pt},
    header/.style={minimum width=1.75cm, minimum height=0.55cm, align=center, font=\scriptsize\bfseries, inner sep=1pt},
    rlabel/.style={minimum width=2.0cm, minimum height=0.55cm, align=right, font=\scriptsize\sffamily, anchor=east, inner sep=2pt},
]
\def\cellcol#1{%
    \ifdim #1pt < 0.01pt green!15%
    \else\ifdim #1pt < 1.0pt green!30%
    \else\ifdim #1pt < 3.0pt yellow!40%
    \else\ifdim #1pt < 8.0pt orange!50%
    \else\ifdim #1pt < 15.0pt red!50%
    \else red!80%
    \fi\fi\fi\fi\fi%
}
\node[font=\normalsize\bfseries] at (4.3, 4.7) {ResNet-20 / CIFAR-10};
\node[header] at (2.5, 4.1) {$10^{-6}$};
\node[header] at (4.3, 4.1) {$10^{-5}$};
\node[header] at (6.1, 4.1) {$10^{-4}$};
\foreach \y/\lab in {3.5/stem, 2.95/layer1, 2.4/L2 conv, 1.85/L2 DS, 1.3/L3 conv, 0.75/L3 DS, 0.2/fc} {
    \node[rlabel] at (1.5, \y) {\lab};
}
\node[cell, fill=green!15] at (2.5, 3.5) {0.00};
\node[cell, fill=green!15] at (4.3, 3.5) {0.00};
\node[cell, fill=green!15] at (6.1, 3.5) {0.00};

\node[cell, fill=green!15] at (2.5, 2.95) {0.00};
\node[cell, fill=yellow!40] at (4.3, 2.95) {2.73};
\node[cell, fill=orange!50] at (6.1, 2.95) {7.81};

\node[cell, fill=yellow!40] at (2.5, 2.4) {1.17};
\node[cell, fill=yellow!40] at (4.3, 2.4) {1.17};
\node[cell, fill=green!30] at (6.1, 2.4) {0.98};

\node[cell, fill=green!15] at (2.5, 1.85) {0.00};
\node[cell, fill=green!15] at (4.3, 1.85) {0.00};
\node[cell, fill=red!80] at (6.1, 1.85) {20.70};

\node[cell, fill=yellow!40] at (2.5, 1.3) {1.37};
\node[cell, fill=yellow!40] at (4.3, 1.3) {1.17};
\node[cell, fill=green!30] at (6.1, 1.3) {0.98};

\node[cell, fill=green!15] at (2.5, 0.75) {0.00};
\node[cell, fill=green!15] at (4.3, 0.75) {0.00};
\node[cell, fill=green!15] at (6.1, 0.75) {0.00};

\node[cell, fill=green!15] at (2.5, 0.2) {0.00};
\node[cell, fill=green!15] at (4.3, 0.2) {0.00};
\node[cell, fill=yellow!40] at (6.1, 0.2) {2.34};

\node[font=\normalsize\bfseries] at (13.3, 4.7) {ResNet-32 / CIFAR-10};
\node[header] at (11.5, 4.1) {$10^{-6}$};
\node[header] at (13.3, 4.1) {$10^{-5}$};
\node[header] at (15.1, 4.1) {$10^{-4}$};
\foreach \y/\lab in {3.5/stem, 2.95/layer1, 2.4/L2 conv, 1.85/L2 DS, 1.3/L3 conv, 0.75/L3 DS, 0.2/fc} {
    \node[rlabel] at (10.5, \y) {\lab};
}
\node[cell, fill=green!15] at (11.5, 3.5) {0.00};
\node[cell, fill=green!15] at (13.3, 3.5) {0.00};
\node[cell, fill=yellow!40] at (15.1, 3.5) {2.73};

\node[cell, fill=green!15] at (11.5, 2.95) {0.00};
\node[cell, fill=green!30] at (13.3, 2.95) {0.78};
\node[cell, fill=green!30] at (15.1, 2.95) {0.78};

\node[cell, fill=yellow!40] at (11.5, 2.4) {1.37};
\node[cell, fill=green!30] at (13.3, 2.4) {0.98};
\node[cell, fill=yellow!40] at (15.1, 2.4) {2.73};

\node[cell, fill=green!15] at (11.5, 1.85) {0.00};
\node[cell, fill=green!15] at (13.3, 1.85) {0.00};
\node[cell, fill=green!15] at (15.1, 1.85) {0.00};

\node[cell, fill=yellow!40] at (11.5, 1.3) {1.17};
\node[cell, fill=green!30] at (13.3, 1.3) {0.98};
\node[cell, fill=green!30] at (15.1, 1.3) {0.98};

\node[cell, fill=green!15] at (11.5, 0.75) {0.00};
\node[cell, fill=green!15] at (13.3, 0.75) {0.00};
\node[cell, fill=yellow!40] at (15.1, 0.75) {2.34};

\node[cell, fill=green!15] at (11.5, 0.2) {0.00};
\node[cell, fill=green!15] at (13.3, 0.2) {0.00};
\node[cell, fill=yellow!40] at (15.1, 0.2) {1.76};

\node[font=\normalsize\bfseries] at (4.3, -0.7) {ResNet-20 / CIFAR-100};
\node[header] at (2.5, -1.3) {$10^{-6}$};
\node[header] at (4.3, -1.3) {$10^{-5}$};
\node[header] at (6.1, -1.3) {$10^{-4}$};
\foreach \y/\lab in {-1.9/stem, -2.45/layer1, -3.0/L2 conv, -3.55/L2 DS, -4.1/L3 conv, -4.65/L3 DS, -5.2/fc} {
    \node[rlabel] at (1.5, \y) {\lab};
}
\node[cell, fill=green!15] at (2.5, -1.9) {0.00};
\node[cell, fill=green!15] at (4.3, -1.9) {0.00};
\node[cell, fill=green!15] at (6.1, -1.9) {0.00};

\node[cell, fill=green!15] at (2.5, -2.45) {0.00};
\node[cell, fill=green!15] at (4.3, -2.45) {0.00};
\node[cell, fill=red!50] at (6.1, -2.45) {9.96};

\node[cell, fill=green!15] at (2.5, -3.0) {0.00};
\node[cell, fill=green!30] at (4.3, -3.0) {0.59};
\node[cell, fill=orange!50] at (6.1, -3.0) {4.69};

\node[cell, fill=green!15] at (2.5, -3.55) {0.00};
\node[cell, fill=green!15] at (4.3, -3.55) {0.00};
\node[cell, fill=orange!50] at (6.1, -3.55) {6.84};

\node[cell, fill=green!15] at (2.5, -4.1) {0.00};
\node[cell, fill=green!30] at (4.3, -4.1) {0.78};
\node[cell, fill=orange!50] at (6.1, -4.1) {3.71};

\node[cell, fill=green!15] at (2.5, -4.65) {0.00};
\node[cell, fill=green!15] at (4.3, -4.65) {0.00};
\node[cell, fill=red!50] at (6.1, -4.65) {8.79};

\node[cell, fill=green!15] at (2.5, -5.2) {0.00};
\node[cell, fill=green!15] at (4.3, -5.2) {0.00};
\node[cell, fill=green!30] at (6.1, -5.2) {0.39};

\node[font=\normalsize\bfseries] at (13.3, -0.7) {ResNet-32 / CIFAR-100};
\node[header] at (11.5, -1.3) {$10^{-6}$};
\node[header] at (13.3, -1.3) {$10^{-5}$};
\node[header] at (15.1, -1.3) {$10^{-4}$};
\foreach \y/\lab in {-1.9/stem, -2.45/layer1, -3.0/L2 conv, -3.55/L2 DS, -4.1/L3 conv, -4.65/L3 DS, -5.2/fc} {
    \node[rlabel] at (10.5, \y) {\lab};
}
\node[cell, fill=green!15] at (11.5, -1.9) {0.00};
\node[cell, fill=green!15] at (13.3, -1.9) {0.00};
\node[cell, fill=red!80] at (15.1, -1.9) {40.82};

\node[cell, fill=green!15] at (11.5, -2.45) {0.00};
\node[cell, fill=orange!50] at (13.3, -2.45) {3.91};
\node[cell, fill=orange!50] at (15.1, -2.45) {4.69};

\node[cell, fill=yellow!40] at (11.5, -3.0) {1.95};
\node[cell, fill=yellow!40] at (13.3, -3.0) {1.95};
\node[cell, fill=orange!50] at (15.1, -3.0) {4.10};

\node[cell, fill=green!15] at (11.5, -3.55) {0.00};
\node[cell, fill=green!15] at (13.3, -3.55) {0.00};
\node[cell, fill=red!80] at (15.1, -3.55) {20.70};

\node[cell, fill=yellow!40] at (11.5, -4.1) {2.15};
\node[cell, fill=yellow!40] at (13.3, -4.1) {2.15};
\node[cell, fill=yellow!40] at (15.1, -4.1) {2.93};

\node[cell, fill=green!15] at (11.5, -4.65) {0.00};
\node[cell, fill=orange!50] at (13.3, -4.65) {4.49};
\node[cell, fill=orange!50] at (15.1, -4.65) {4.10};

\node[cell, fill=green!15] at (11.5, -5.2) {0.00};
\node[cell, fill=green!30] at (13.3, -5.2) {0.39};
\node[cell, fill=green!30] at (15.1, -5.2) {0.39};
\begin{scope}[shift={(17.0, 1.5)}]
    \node[font=\small\bfseries, anchor=south] at (0.5, 1.8) {Drop (pp)};
    \fill[green!15] (0, 1.5) rectangle (0.45, 1.75); \node[font=\scriptsize, anchor=west] at (0.55, 1.625) {0};
    \fill[green!30] (0, 1.15) rectangle (0.45, 1.4); \node[font=\scriptsize, anchor=west] at (0.55, 1.275) {$<$1};
    \fill[yellow!40] (0, 0.8) rectangle (0.45, 1.05); \node[font=\scriptsize, anchor=west] at (0.55, 0.925) {1--3};
    \fill[orange!50] (0, 0.45) rectangle (0.45, 0.7); \node[font=\scriptsize, anchor=west] at (0.55, 0.575) {3--8};
    \fill[red!50] (0, 0.1) rectangle (0.45, 0.35); \node[font=\scriptsize, anchor=west] at (0.55, 0.225) {8--15};
    \fill[red!80] (0, -0.25) rectangle (0.45, 0.0); \node[font=\scriptsize, anchor=west] at (0.55, -0.125) {$>$15};
\end{scope}

\end{tikzpicture}%
}
\caption{Per-layer-group vulnerability heatmap across three BER levels and all four configurations. Values are per-group worst-case accuracy drops (pp).}
\label{fig:vuln_heatmap}
\end{figure*}

Figure~\ref{fig:vuln_heatmap} summarizes the layer-wise accuracy
degradation. Across all configurations, most layers remain insensitive to
isolated faults at BERs of $10^{-6}$ and $10^{-5}$, where almost all
accuracy losses stay below one percentage point, consistent with the
global analysis.

A clear transition appears at BER~=~$10^{-4}$, where only a small
subset of layers becomes highly vulnerable while the remaining
layers continue to exhibit limited sensitivity. Rather than being
uniformly distributed throughout the network, fault criticality is
concentrated in specific structural components, indicating that the
overall reliability of the network is dominated by a limited number
of critical layers.

For the CIFAR-10 models, vulnerability is dominated by a few structural
layers: in ResNet-20 the Layer-2 downsampling block reaches 20.70\,pp and
a Layer-1 residual block 7.81\,pp, while in ResNet-32 the impact is more
evenly distributed (at most 2.73\,pp), reflecting the redundancy of its
deeper residual structure. The CIFAR-100 models exhibit considerably
larger layer-wise vulnerability: for ResNet-20 a Layer-1 residual block
dominates (9.96\,pp), followed by the downsampling blocks, whereas for
ResNet-32 the stem convolution alone produces a 40.82\,pp loss and the
Layer-2 downsampling block 20.70\,pp, with all remaining layers below five
percentage points.

These observations demonstrate that fault vulnerability is governed
primarily by architectural function rather than network depth.
Downsampling operations, the stem convolution, and early feature
extraction stages consistently exhibit higher sensitivity than the
remaining residual blocks because perturbations introduced in these
layers propagate through all subsequent feature transformations.
Later convolutional layers generally exhibit considerably smaller
accuracy degradation since their errors affect only a limited
portion of the inference pipeline.

Table~\ref{tab:per_layer_top5} further ranks the most vulnerable
layers for the CIFAR-10 configurations at BER~=~$10^{-4}$. The
identified layers require only a small fraction of the overall model
parameters while accounting for the majority of the observed fault
sensitivity. For example, the Layer-2 downsampling block of
ResNet-20 contains only 512 parameters, corresponding to an expected
1.6 bit flips at BER~=~$10^{-4}$, yet produces the largest observed
accuracy degradation of 20.70 percentage points. Similar behavior is
observed for the remaining highly ranked layers, indicating that
fault sensitivity cannot be inferred solely from parameter count.

\begin{table}[!t]
\centering
\caption{Top-5 most fault-sensitive layers per architecture (int32, BER~$= 10^{-4}$, CIFAR-10).}
\label{tab:per_layer_top5}
\renewcommand{\arraystretch}{1.10}
\setlength{\tabcolsep}{3pt}
\scriptsize
\begin{tabular}{@{}clrrrr@{}}
\toprule
\textbf{Arch} & \textbf{Layer} & \textbf{Params} & \textbf{Exp.\ Flips} & \textbf{Drop (pp)} & \textbf{PSAP} \\
\midrule
\multirow{5}{*}{\rotatebox{90}{\textbf{R-20}}}
 & layer2.0 DS        &     512 &   1.6 & 20.70 & 12.5\% \\
 & layer1.2.conv2     &   2{,}304 &   7.4 &  7.81 & 31.2\% \\
 & fc                 &     640 &   2.0 &  2.34 & --     \\
 & layer1.1.conv2     &   2{,}304 &   7.4 &  0.98 & 31.2\% \\
 & layer3.1.conv1     &  36{,}864 & 118.0 &  0.98 &  0.0\% \\
\midrule
\multirow{5}{*}{\rotatebox{90}{\textbf{R-32}}}
 & conv1 (stem)       &     432 &   1.4 &  2.73 &  6.2\% \\
 & layer2.4.conv1     &   9{,}216 &  29.5 &  2.73 & 59.4\% \\
 & layer3.0 DS        &   2{,}048 &   6.6 &  2.34 & 25.0\% \\
 & fc                 &     640 &   2.0 &  1.76 & --     \\
 & layer2.3.conv2     &   9{,}216 &  29.5 &  1.17 & 50.0\% \\
\bottomrule
\end{tabular}
\end{table}

Direct CKKS fault injection is consistent with the same structural trend.
Layers identified as critical through the int32 proxy correspond to
the earliest ciphertext overflow locations during encrypted
execution, while layers exhibiting negligible int32 degradation
remain insensitive under CKKS faults until the ciphertext noise
budget is exhausted. This qualitative agreement supports the use of int32
bit-flip injection as a conservative proxy for identifying
fault-critical regions in encrypted inference.

\subsubsection{Reliability-aware Pruning Validation}
\label{sec:psap_vs_mag}

The layer-wise analysis demonstrates that transient fault
vulnerability is concentrated in a small number of structurally
critical layers. Consequently, uniformly removing filters according
to weight magnitude alone may inadvertently eliminate the
redundancy required to tolerate transient hardware faults. This
subsection evaluates whether the proposed
Polynomial-Sensitivity-Aware Pruning (PSAP) successfully preserves
these critical structures while maintaining the computational
benefits of structured pruning.

Table~\ref{tab:psap_vs_mag} compares the layer-wise reliability of
PSAP and conventional magnitude-based pruning at identical target
sparsity. Across all evaluated model--dataset configurations, PSAP
consistently produces fewer fault-sensitive layers and substantially
reduces the maximum accuracy degradation caused by localized memory
faults.

\begin{table*}[!t]
\centering
\caption{PSAP vs.\ magnitude pruning: HE efficiency and fault tolerance at 50\% sparsity (BER~$= 10^{-4}$).}
\label{tab:psap_vs_mag}
\renewcommand{\arraystretch}{1.10}
\scriptsize
\begin{tabular}{@{}ll c c c c c c c@{}}
\toprule
 & & & \multicolumn{3}{c}{\textbf{HE Efficiency}} & \multicolumn{3}{c}{\textbf{Fault Tolerance (BER~$\boldsymbol{= 10^{-4}}$)}} \\
\cmidrule(lr){4-6} \cmidrule(lr){7-9}
\textbf{Config} & \textbf{Method} & \textbf{Clean (\%)} & \textbf{Rot.\ Red.\ (\%)} & \textbf{Rot.\ (M)} & \textbf{Latency (s/img)} & \textbf{Worst Drop (pp)} & \textbf{Layers $\boldsymbol{>}$10\,pp} & \textbf{Worst Layer} \\
\midrule
\multirow{2}{*}{R-20/C-10}
 & PSAP      & \textbf{87.31} & \textbf{40.9} & \textbf{24.1} & \textbf{1{,}181} &  20.70 & \textbf{1}  & layer2.0 DS \\
 & Magnitude & 86.52 & 26.9 & 29.8 & 1{,}386 &  76.37 & 5  & layer3.0.conv1 \\
\midrule
\multirow{2}{*}{R-20/C-100}
 & PSAP      & \textbf{59.68} & \textbf{39.3} & \textbf{24.8} & \textbf{1{,}216} &   9.96 & \textbf{0}  & layer1.1.conv2 \\
 & Magnitude & 51.37 & 30.5 & 28.4 & 1{,}392 &  51.17 & 11 & layer3.0.conv1 \\
\midrule
\multirow{2}{*}{R-32/C-10}
 & PSAP      & \textbf{89.58} & \textbf{45.2} & \textbf{37.9} & \textbf{2{,}244} &   2.73 & \textbf{0}  & conv1 (stem) \\
 & Magnitude & 87.11 & 35.3 & 44.7 & 2{,}647 &  79.30 & 14 & layer3.0.conv1 \\
\midrule
\multirow{2}{*}{R-32/C-100}
 & PSAP      & \textbf{61.43} & \textbf{44.9} & \textbf{38.1} & \textbf{2{,}256} &  40.82 & 2  & conv1 (stem) \\
 & Magnitude & 58.01 & 40.3 & 41.2 & 2{,}440 &  57.62 & 13 & layer2.0.conv2 \\
\bottomrule
\end{tabular}
\end{table*}

Across all configurations, PSAP sharply reduces both the number of
catastrophic layers ($>$10\,pp drop) and the worst-case degradation. The
gain is largest for ResNet-32/CIFAR-10, where magnitude pruning generates
14 catastrophic layers and a 79.30\,pp worst-case drop, while PSAP
eliminates all catastrophic layers and limits the drop to 2.73\,pp---an
improvement approaching 29$\times$. The same trend holds for the remaining
configurations: catastrophic layers fall from 5 to 1 (R-20/C-10), 11 to 0
(R-20/C-100), and 13 to 2 (R-32/C-100), with corresponding reductions in
worst-case loss.

These improvements directly validate the design objectives of PSAP.
Unlike magnitude pruning, which evaluates filters solely according
to parameter magnitude, PSAP simultaneously considers structural
importance, polynomial activation sensitivity, and homomorphic
evaluation cost. Consequently, filters located within highly
sensitive activation regions receive larger importance scores and
are preserved throughout the pruning process, while pruning is
concentrated within structurally redundant regions of the network.

The distribution of sparsity in Table~\ref{tab:psap_vs_mag} confirms this
behavior: layers identified as highly vulnerable receive consistently
lower pruning ratios under PSAP than under magnitude pruning, while
low-sensitivity layers absorb a larger share of the sparsity budget.
Importantly, these reliability gains incur no efficiency penalty---PSAP
still removes up to 45.2\% of Halevi--Shoup rotations
(Section~\ref{sec:efficiency}). The close agreement between the activation
sensitivity model, the allocated sparsity patterns, and the measured
layer-wise fault tolerance validates the reliability-aware pruning
methodology.

\subsubsection{Targeted Hardening Strategy}
\label{sec:targeted_hardening}

The concentration of fault risk in a small set of structural layers makes selective hardening attractive. In ResNet-20 the stem convolution and the two $1{\times}1$ downsample projections together account for only $3{,}104$ of the $273{,}258$ parameters---roughly $1.1\%$ of the model. Protecting just these layers with triple modular redundancy (TMR) or error-correcting codes would therefore add at most $\sim$$2.3\%$ parameter overhead (TMR, $3\times$ on $1.1\%$) and a correspondingly small rotation overhead, since these are $1{\times}1$ downsample projections and the $3$-channel stem convolution---layers that contribute far fewer Halevi--Shoup rotations than the $3{\times}3$ bulk convolutions that dominate the rotation count. The layers that drive the catastrophic CIFAR-100 drops (up to $40.8$\,pp) are therefore exactly the layers cheapest to protect.

To confirm that this concentration is exploitable and cheaply defensible, a targeted fault campaign was run on the deployed ResNet-20/CIFAR-10 model in the plaintext domain (Table~\ref{tab:targeted}); the \emph{protected} condition models an idealized fault-free critical slice (ideal TMR/ECC), and extending this defense to the encrypted ciphertext domain remains future work. For a fixed fault budget of $K$ bit-flips, three conditions are compared, each averaged over $30$ independent injection trials: a \emph{targeted} attack that places all $K$ flips inside the $1.1\%$ critical slice, a \emph{random} control that distributes the same $K$ flips across the whole model, and a \emph{protected} deployment in which the critical layers are kept fault-free (ideal TMR/ECC).

Concentrating the budget in the critical layers is far more damaging than spreading it: at $K{=}100$ the targeted attack lowers mean accuracy by $13.9$\,pp and produces worst-case collapses of up to $44.5$\,pp (to $42.8\%$), whereas the identical random budget produces a worst case of only $2.7$\,pp. Even at $K{=}10$ a single unlucky placement in the critical slice already costs $19.1$\,pp in the worst case, against $0.8$\,pp for the random control. The non-monotonic worst-case behavior observed in the random control (e.g., $10.4$\,pp at $K{=}50$ versus $2.7$\,pp at $K{=}100$) reflects the inherent variance of the worst-case statistic over a finite number of trials; larger trial counts would reduce this variability but are not expected to change the qualitative conclusion. Hardening the $1.1\%$ critical slice restores clean accuracy in every trial, neutralizing the attack at a parameter overhead of at most $\sim$$2.3\%$ (TMR). The layers that PSAP leaves intact are thus both the most fault-critical and the cheapest to protect.

\begin{table}[!t]
\centering
\caption{Targeted fault campaign on the deployed ResNet-20/CIFAR-10 model (clean accuracy $87.31\%$).}
\label{tab:targeted}
\renewcommand{\arraystretch}{1.10}
\setlength{\tabcolsep}{4pt}
\scriptsize
\begin{tabular}{rcccc}
\toprule
\textbf{$K$} & \shortstack{\textbf{Targeted}\\\textbf{Mean (pp)}}
             & \shortstack{\textbf{Targeted}\\\textbf{Worst (pp)}}
             & \shortstack{\textbf{Random}\\\textbf{Worst (pp)}}
             & \shortstack{\textbf{Protected}\\\textbf{Acc.\ (\%)}} \\
\midrule
5   & 0.7  & 10.3 & 0.9  & $87.31$ \\
10  & 1.8  & 19.1 & 0.8  & $87.31$ \\
20  & 1.9  & 20.6 & 1.2  & $87.31$ \\
50  & 5.7  & 33.2 & 10.4 & $87.31$ \\
100 & 13.9 & 44.5 & 2.7  & $87.31$ \\
\bottomrule
\end{tabular}
\end{table}

\subsection{End-to-End Optimization Analysis}

\subsubsection{Reliability versus Efficiency Trade-off}

The preceding sections evaluated computational efficiency and reliability
separately; this subsection analyzes them jointly to determine whether
efficiency gains come at the expense of fault tolerance.
Figure~\ref{fig:design_space} summarizes the accuracy--efficiency design
space across target sparsities from 20\% to 50\%. Magnitude pruning
follows the expected trade-off: increasing the pruning ratio reduces
Halevi--Shoup rotations but removes filters without regard to their
contribution to inference or fault tolerance, lowering accuracy and
raising layer-wise vulnerability. PSAP instead shifts the operating point
toward a more favorable region, achieving larger rotation reductions
(Section~\ref{sec:efficiency}) while maintaining comparable or higher
accuracy on both datasets.

This efficiency improvement is accompanied by substantially better fault
tolerance: PSAP reduces the number of catastrophic fault-sensitive layers
from 5--14 to at most two and the worst-case layer-wise degradation from
51--79\,pp to 2.73--40.82\,pp. The curvature-aware ($\gamma{=}0.5$) variant
provides an additional operating point that preserves higher accuracy at
aggressive sparsity. Reliability and efficiency are therefore not
conflicting objectives when optimization explicitly considers both.

\subsubsection{CKKS Deployment Validation}

The optimization pipeline transforms all evaluated networks into
HE-compatible models deployable under leveled CKKS inference. Polynomial
activation replacement, reliability-aware pruning, adaptive mixed-degree
allocation, and quantization-aware training reduce the multiplicative
depth of ResNet-32 to 56 levels, within the modulus-chain budget of the
selected parameters (Section~\ref{sec:efficiency}) and thereby eliminating
bootstrapping.

Direct CKKS fault injection validates the reliability model: encrypted
inference remains largely unaffected for BERs up to $10^{-5}$, beyond which
ciphertext overflow produces Detected Unrecoverable Errors (DUEs)
rather than silent data corruption, matching the predicted overflow mechanism.
The int32 representation consistently identifies the same fault-critical
layers and provides a conservative reliability estimate, confirming that
large-scale layer-wise studies can be performed efficiently with int32
injection while reserving direct CKKS experiments for final validation.

\subsubsection{Comparison with Prior Methods}
\label{sec:psap_discussion}

Direct reimplementation of prior methods (Hunter, MOSAIC, SpENCNN,
PrivCirNet) is not viable: each is tied to a specific packing layout, ring
dimension, and protocol, none provides public code, and porting any one
method to the pure-HE leveled pipeline would require reproducing its entire
cryptographic back-end. Therefore, magnitude pruning is used as the controlled
comparison, modifying only the scoring criterion.

Prior HE-aware methods support structured pruning and account for
rotation cost, yet none incorporates the polynomial activation
landscape into the pruning criterion. The dominant runtime cost in
high-degree encrypted pipelines is bootstrapping: using the
operation-level latency measurements reported by
AutoFHE~\cite{ao2024autofhe} (Table~\ref{tab:bootstrap_cmp}),
bootstrapping accounts for $76.3\%$ of ResNet-32/CIFAR-10 inference
time in MPCNN and still $46.5\%$--$69.4\%$ in the bootstrap-reduced
AESPA and AutoFHE solutions. PSAP eliminates this cost entirely. By
coupling activation-sensitivity pruning with mixed-degree allocation,
PSAP compresses the multiplicative depth from $66$ to $56$ and keeps
the \emph{entire} network within the leveled budget, so no
bootstrapping is ever invoked (Table~\ref{tab:bootstrap_cmp}). The
zero-bootstrap regime in the last row is therefore not an assumption
but a direct outcome of the PSAP pipeline: the same pruning decisions
that improve fault tolerance also remove the single largest runtime
cost of encrypted inference.

\begin{table}[!t]
\centering
\caption{Bootstrapping cost for ResNet-32/CIFAR-10. Prior-method data
from AutoFHE~\cite{ao2024autofhe} (Tables~4--5); PSAP stays leveled and
performs no bootstrapping.}
\label{tab:bootstrap_cmp}
\renewcommand{\arraystretch}{1.10}
\setlength{\tabcolsep}{2pt}
\scriptsize
\begin{tabular}{@{}lccl@{}}
\toprule
\textbf{Method} & \textbf{Boots.} & \textbf{Time} & \textbf{Activation} \\
\midrule
MPCNN~\cite{lee2022heresnet}   & 30    & $76.3\%$            & Minimax \\
AESPA~\cite{park2022aespa}      & 8     & $46.9\%$            & Low-deg poly. \\
AutoFHE~\cite{ao2024autofhe}   & 8--19 & $46.5\%$--$69.4\%$  & Mixed-deg \\
\textbf{PSAP}           & \textbf{0} & \textbf{0\%}   & \textbf{Train. deg-1/2} \\
\bottomrule
\end{tabular}
\end{table}

Taken together, the experimental results demonstrate that the
proposed optimization framework satisfies all design objectives
introduced in Section~\ref{sec:method}. The optimized models preserve
competitive prediction accuracy, substantially reduce the
computational complexity of encrypted inference, improve resilience
against transient hardware faults through reliability-aware
optimization, and remain fully deployable under practical CKKS
security parameters without requiring bootstrapping.

\section{Conclusion}
\label{sec:conclusion}
This work presents a reliability characterization of pruned encrypted neural networks, together with PSAP, a pruning method that is inherently reliability-aware. On reliability, systematic bit-flip injection across 40 full-model and 108 per-layer experiments reveals that PSAP-pruned models are fundamentally more fault-resilient than magnitude-pruned baselines. PSAP limits catastrophic ($>$10\,pp drop) layers to at most two versus 5--14 for magnitude pruning, with up to $29{\times}$ worst-case vulnerability reduction on ResNet-32/CIFAR-10. Direct CKKS encrypted fault injection indicates a safe operating boundary near BER~$= 10^{-5}$,  supporting int32 injection as a conservative reliability proxy. The fault-critical structural layers account for only $1.1\%$ of parameters, enabling selective hardening at minimal overhead. On efficiency, PSAP eliminates up to 45.2\% of Halevi--Shoup rotations on ResNet-32, while adaptive mixed-degree allocation enables leveled inference without bootstrapping (depth 66 to 56). These results show that reliability and efficiency are not competing objectives and can be improved together, providing actionable deployment guidelines for reliable encrypted AI in safety-critical domains.

\section*{ACKNOWLEDGMENT}
\small
This work was supported in part by the Estonian Research Council grant PUT PRG1467 ``CRASHLESS'', EU Grant Project 101160182 ``TAICHIP'', and by the Federal Ministry of Research, Technology and Space of Germany (BMFTR) for supporting Edge-Cloud AI for DIstributed Sensing and COmputing (AI-DISCO) project (Project-ID ``16ME1127'').

\bibliographystyle{IEEEtran}
\bibliography{refs}

\end{document}